\documentclass[reprint,aps,prb,twocolumn,amsmath,amssymb,showpacs,preprintnumbers]{revtex4-2}

\usepackage{graphicx}[]
\usepackage{dcolumn}
\usepackage{bm}
\usepackage{multirow}

\usepackage{braket}
\usepackage{color}
\usepackage{ulem}
\usepackage{url}
\usepackage{hyperref}

\begin{document}

\title{
Checkerboard bubble lattice formed by octuple-period quadruple-$Q$ spin density waves
}

\author{Satoru Hayami}
\affiliation{
Graduate School of Science, Hokkaido University, Sapporo 060-0810, Japan
}

\begin{abstract}
We investigate multiple-$Q$ instability on a square lattice at particular ordering wave vectors. 
We find that a superposition of quadruple-$Q$ spin density waves, which are connected by fourfold rotational and mirror symmetries, gives rise to a checkerboard bubble lattice with a collinear spin texture as a result of the geometry among the constituent ordering wave vectors in the Brillouin zone. 
By performing the simulated annealing for a fundamental spin model, we show that such a checkerboard bubble lattice is stabilized under an infinitesimally small easy-axis two-spin anisotropic interaction and biquadratic interaction at zero field, while it is degenerate with an anisotropic double-$Q$ state in the absence of the biquadratic interaction. 
The obtained multiple-$Q$ structures have no intensities at high-harmonic wave vectors in contrast to other multiple-$Q$ states, such as a magnetic skyrmion lattice. 
We also show that the checkerboard bubble lattice accompanies the charge density wave and exhibits a nearly flat band dispersion in the electronic structure.  
Our results provide another route to realize exotic multiple-$Q$ spin textures by focusing on the geometry and symmetry in terms of the wave vectors in momentum space. 
\end{abstract}

\maketitle

\section{Introduction}

A multiple-$Q$ state, which corresponds to a superposition of multiple spin density waves, has drawn considerable interest in condensed matter physics, since it manifests itself in unusual magnetism but also in unconventional transport and cross-correlated response. 
One of the most familiar examples is a magnetic skyrmion lattice (SkL)~\cite{rossler2006spontaneous, Muhlbauer_2009skyrmion, yu2010real, yu2011near, nagaosa2013topological}, which ubiquitously appears in various lattice structures as a different multiple-$Q$ superposition~\cite{Tokura_doi:10.1021/acs.chemrev.0c00297}: double-$Q$ SkL in the tetragonal lattice structure~\cite{Yi_PhysRevB.80.054416, khanh2020nanometric, Karube_PhysRevB.102.064408, khanh2022zoology}, triple-$Q$ SkL in the hexagonal lattice structure~\cite{Muhlbauer_2009skyrmion, yu2010real, Seki_PhysRevB.85.220406}, and sextuple-$Q$ SkL in the cubic lattice structure~\cite{Hayami_PhysRevB.107.174435}.  
Another example is a magnetic hedgehog consisting of a periodic structure of monopole and antimonopole, which has been found as a consequence of the triple-$Q$ or quadruple-$Q$ spin density waves~\cite{Ishiwata_PhysRevB.84.054427, tanigaki2015real, kanazawa2017noncentrosymmetric, Ishiwata_PhysRevB.84.054427, fujishiro2019topological, Ishiwata_PhysRevB.101.134406, Rogge_PhysRevMaterials.3.084404}. 
Moreover, various multiple-$Q$ spin structures have been suggested/proposed from both theoretical and experimental studies, such as a vortex lattice~\cite{Kamiya_PhysRevX.4.011023, Liu_PhysRevB.94.174424, Seabra_PhysRevB.93.085132, Hayami_PhysRevB.94.174420, Chern_PhysRevB.95.144427, takagi2018multiple, hayami2021phase, Kobayashi_PhysRevB.106.L140406, Zhang_PhysRevB.105.024411, Ishitobi_PhysRevB.107.104413}, a chiral stripe~\cite{Solenov_PhysRevLett.108.096403, Ozawa_doi:10.7566/JPSJ.85.103703, khanh2022zoology, Wood_PhysRevB.107.L180402, zager2023double}, and a ripple state~\cite{Shimokawa_PhysRevLett.123.057202}. 
As a common feature in these multiple-$Q$ states, the constituent ordering wave vectors are connected to each other by the rotational symmetry of the lattice structures.

\begin{figure}[t!]
\begin{center}
\includegraphics[width=1.0 \hsize ]{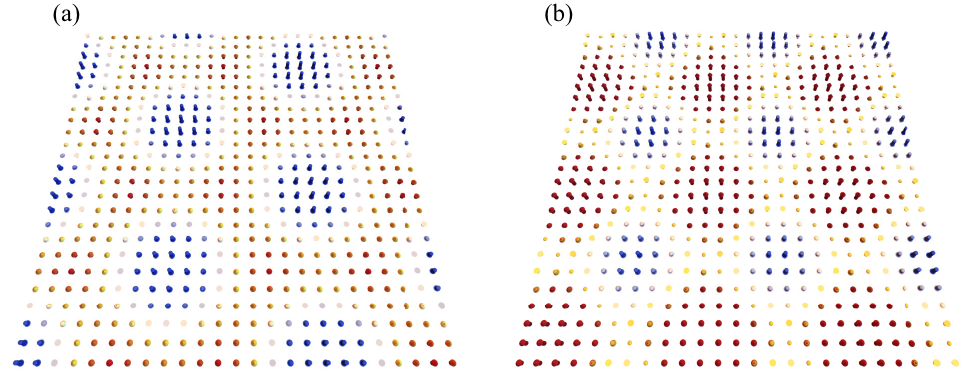} 
\caption{
\label{fig: ponti}
Schematic pictures of (a) the triangular bubble lattice modulated by the triple-$Q$ spin density waves and (b) the square bubble lattice modulated by the double-$Q$ spin density waves. 
The arrows and the color represent the spin moments and its $z$ direction component; the blue, red, and yellow stand for the up, down, and zero moments. 
}
\end{center}
\end{figure}

We here investigate a different type of a multiple-$Q$ state, which is characterized by a superposition of collinear spin textures, i.e., a bubble lattice~\cite{lin1973bubble, Garel_PhysRevB.26.325, takao1983study}. 
Although such a bubble lattice does not exhibit a topological Hall effect in contrast to the SkL, it can be potentially applied to the growing field of antiferromagnetic/ferrimagnetic spintronics~\cite{jungwirth2016antiferromagnetic, caretta2018fast}, since unconventional transport property is expected~\cite{callen1971dynamics, de1980dynamic, Moutafis_PhysRevB.79.224429}. 
The isolated bubble and bubble lattices have been so far observed in easy-axis magnets~\cite{xiao2020spin}, such as Fe/Rh atomic bilayers on the Ir(111) surface~\cite{gutzeit2022nano} hosting the triangular bubble lattice in Fig.~\ref{fig: ponti}(a) and CeAuSb$_2$~\cite{Marcus_PhysRevLett.120.097201, Park_PhysRevB.98.024426, Seo_PhysRevX.10.011035, seo2021spin} hosting the square bubble lattice in Fig.~\ref{fig: ponti}(b); the former is described by the triple-$Q$ collinear state, while the latter is described by the double-$Q$ collinear state. 
Simultaneously, theoretical model calculations clarified various mechanisms of the bubble lattices, such as the competing exchange interaction with single-ion magnetic anisotropy~\cite{Hayami_PhysRevB.93.184413}, biquadratic interaction~\cite{hayami2020multiple}, high-harmonic wave-vector interaction~\cite{hayami2023magnetic}, and thermal fluctuations~\cite{Hayami_10.1088/1367-2630/ac3683}. 

In the present study, we propose a further intriguing bubble lattice in tetragonal magnets, which is characterized by a different multiple-$Q$ superposition from the conventional bubble lattices. 
By focusing on the geometry and symmetry of the constituent ordering wave vectors, we find that a checkerboard bubble lattice is engineered on a square lattice by superposing quadruple-$Q$ spin density waves with the octuple-period ordering wave vector, where we call it 4$Q$ CBL. 
We show that the 4$Q$ CBL state is stabilized by taking into account an infinitesimally small easy-axis two-spin anisotropic interaction and positive biquadratic interaction by performing analytical and numerical analyses. 
In contrast to the conventional SkL and bubble lattice, the 4$Q$ CBL state does not have the intensity at high-harmonic ordering wave vectors, which makes it difficult to distinguish it from a multi-domain single-$Q$ state by small-angle neutron scattering experiments. 
Instead, the real-space observation through spectroscopic-imaging scanning tunneling microscopy measurements~\cite{Yasui2020imaging} is one of the probes to identify the 4$Q$ CBL state. 
Moreover, the 4$Q$ CBL state exhibits a nearly flat band structure in the strong-coupling regime. 
The present result indicates that multiple-$Q$ superpositions can give rise to further exotic spin textures, which have not been observed in both theory and experiment, depending on the symmetry and geometry of the ordering wave vectors. 

The organization of this paper is as follows. 
In Sec.~\ref{sec: Model and method}, we introduce a minimal spin model to induce the instability toward the 4$Q$ CBL state. 
We also outline the numerical method based on the simulated annealing. 
Then, we show how to construct the 4$Q$ CBL state by the multiple-$Q$ superposition of the octuple-period spin density waves in Sec.~\ref{sec: Results}. 
Furthermore, we discuss the stability region of the 4$Q$ CBL state while changing an external magnetic field and easy-axis two-spin magnetic anisotropy. 
In Sec.~\ref{sec: Discussion}, we discuss the charge density waves and the electronic band structure in the presence of the 4$Q$ CBL spin texture. 
Section~\ref{sec: Summary} summarizes this paper.

\section{Model and method}
\label{sec: Model and method}

We consider an effective spin model on a two-dimensional square lattice under the space group $P4/mmm$, which is obtained by tracing out the itinerant electron degree of freedom in the Kondo lattice model with the classical spin~\cite{hayami2021topological}. 
By supposing a weak Kondo coupling compared to the bandwidth of the itinerant electrons, one can obtain the following spin model based on the perturbative expansion~\cite{Hayami_PhysRevB.95.224424}: 
\begin{align}
\label{eq: Ham_pre}
\mathcal{H}=  &
- \sum_{\bm{q}}J_{\bm{q}} (\bm{S}_{\bm{q}}\cdot \bm{S}_{-\bm{q}} +I^z S^{z}_{\bm{q}}S^{z}_{-\bm{q}}) \nonumber \\
&+\sum_{\bm{q}}\frac{K_{\bm{q}}}{N} (\bm{S}_{\bm{q}} \cdot  \bm{S}_{-\bm{q}} +I^z S^{z}_{\bm{q}}S^{z}_{-\bm{q}})^2  
-  H \sum_{i}   S^z_{i},
\end{align}
where $\bm{S}_{\bm{q}}=(S^x_{\bm{q}}, S^y_{\bm{q}}, S^z_{\bm{q}})$ is the $\bm{q}$ component of the spin moment; $S^\eta_{\bm{q}}$ is related to the classical spin $S^\eta_i$ at site $i$ via the Fourier transformation. 
We fix the spin length at each site as $|\bm{S}_i|=1$ and take the lattice constant as unity. 
The first term represents the bilinear exchange interaction, which is obtained from the lowest-order contribution in the perturbative expansion; the positive coupling constant $J_{\bm{q}}>0$ is proportional to $J^2_{\rm K}$, where $J_{\rm K}$ stands for the Kondo coupling between the itinerant electron spin and localized spin. 
This bilinear term is referred to as the Ruderman-Kittel-Kasuya-Yosida (RKKY) interaction~\cite{Ruderman, Kasuya, Yosida1957}. 
We suppose the anisotropic form factor for the $xy$ and $z$ components, which originates from the relativistic spin--orbit coupling; $I^z>0$ ($I^z<0$) represents the easy-axis (easy-plane) anisotropy. 
Meanwhile, we ignore other magnetic anisotropy allowed from the symmetry of the square lattice, such as the bond-dependent magnetic anisotropy~\cite{Yambe_PhysRevB.106.174437}, since it does not affect the stability of the 4$Q$ CBL state. 
The second term denotes the biquadratic interaction with the positive coupling constant $K_{\bm{q}}>0$, which is derived from the second-lowest-order contribution in the perturbative expansion, i.e., $K_{\bm{q}} \propto J^4_{\rm K}$; $N$ stands for the number of spins in the system. 
We neglect other multiple spin interactions in the form of $(\bm{S}_{\bm{q}_1}\cdot \bm{S}_{\bm{q}_2})(\bm{S}_{\bm{q}_3}\cdot \bm{S}_{\bm{q}_4})$ with $\bm{q}_1+\bm{q}_2+\bm{q}_3+ \bm{q}_4=\bm{0}$ by supposing the strong nesting of the Fermi surface in the band structure~\cite{Akagi_PhysRevLett.108.096401, Hayami_PhysRevB.90.060402, Hayami_PhysRevB.95.224424}. 
Since $K_{\bm{q}}$ corresponds to the higher-order term than $J_{\bm{q}}$ in terms of the Kondo coupling, we set $J_{\bm{q}} \gg K_{\bm{q}}$. 
The third term represents the Zeeman coupling in the presence of an external magnetic field along the out-of-plane direction. 

\begin{figure}[t!]
\begin{center}
\includegraphics[width=1.0 \hsize ]{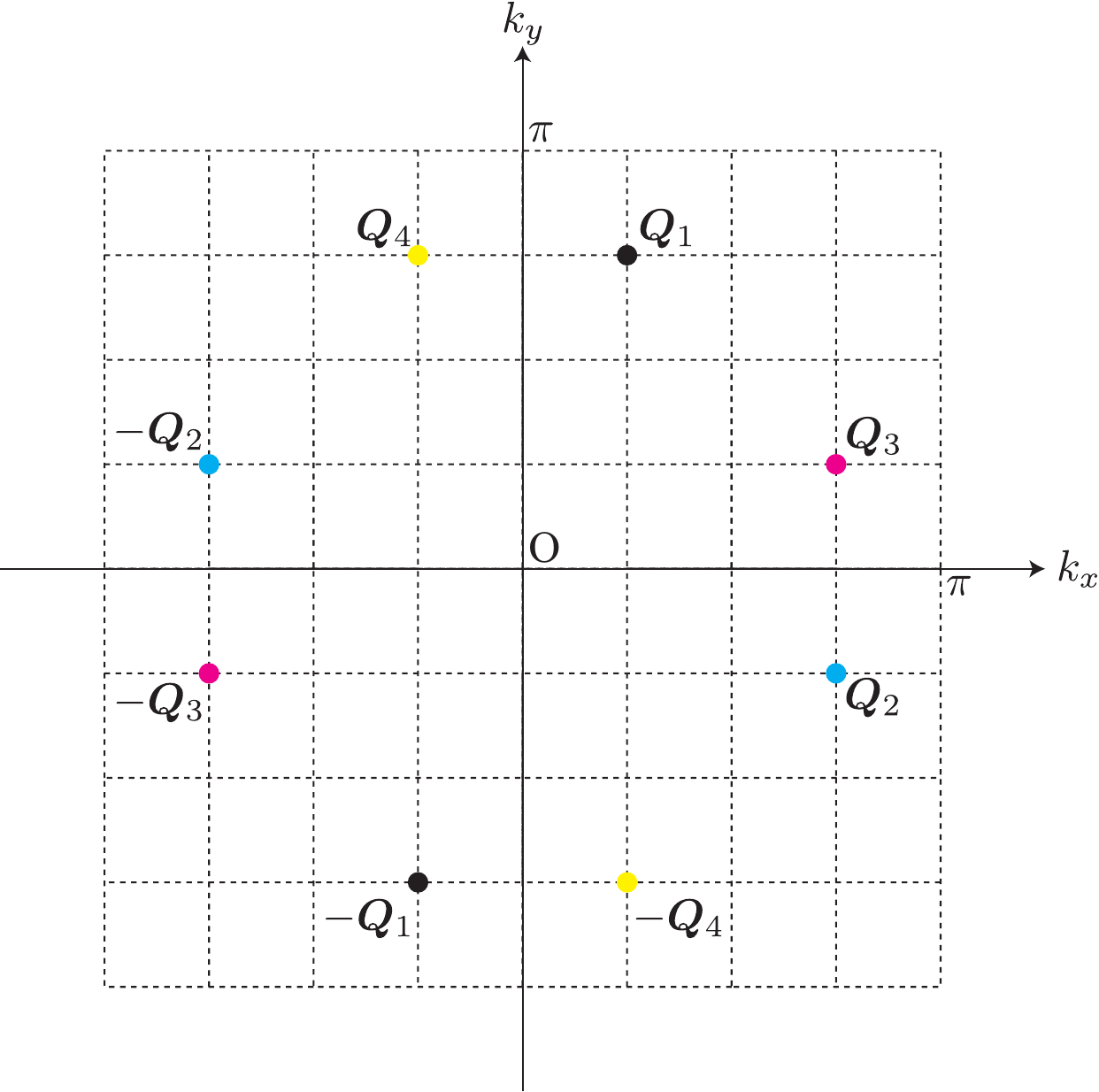} 
\caption{
\label{fig: Qvec}
Ordering wave vectors $\bm{Q}_1$--$\bm{Q}_4$ in the Brillouin zone with $-\pi < k_x \leq \pi$ and $-\pi < k_y \leq \pi$. 
$\bm{Q}_1$ and $\bm{Q}_2$ ($\bm{Q}_3$ and $\bm{Q}_4$) are connected by the fourfold rotational symmetry. 
$\bm{Q}_1$ and $\bm{Q}_3$ ($\bm{Q}_2$ and $-\bm{Q}_4$) are connected by the mirror symmetry along the [110] ([$\bar{1}10$]) axis. 
$\bm{Q}_1$ and $\bm{Q}_4$ ($\bm{Q}_2$ and $\bm{Q}_3$) are connected by the mirror symmetry along the $k_y$ ($k_x$) axis. 
}
\end{center}
\end{figure}

We investigate the ground-state phase diagram of the model in Eq.~(\ref{eq: Ham_pre}) by changing the model parameters $J_{\bm{q}}$, $K_{\bm{q}}$, $I^z$, and $H$. 
Since we suppose $J_{\bm{q}} \gg K_{\bm{q}}$, the magnetic instability at zero temperature occurs at $\bm{q}^*$ satisfying $J_{\bm{q}^*} > J_{\bm{q}'}$ for $\bm{q}^* \neq \bm{q}'$; the circular (elliptical) spiral state with the spiral pitch $\bm{q}^*$ is expected for $I^z=0$ ($I^z>0$). 
In this situation, almost all of the wave vectors $\bm{q}'$ do not contribute to the energy, which indicates that the interactions in the $\bm{q}'$ channel are not relevant to determine the ground-state phase diagram. 
Similar attempts are done for investigating the square SkL, where several mechanisms have been clarified by analyzing minimal models, such as the Dzyaloshinskii-Moriya interaction~\cite{Yi_PhysRevB.80.054416, Hayami_PhysRevLett.121.137202, hayami2022square}, positive biquadratic interaction~\cite{Hayami_PhysRevB.103.024439, hayami2023widely}, magnetic anisotropy~\cite{Wang_PhysRevB.103.104408, Hayami_PhysRevB.105.174437}, dipolar interaction~\cite{Utesov_PhysRevB.103.064414}, and high-harmonic wave-vector interaction~\cite{hayami2022multiple, Hayami_PhysRevB.105.104428, hayami2023widely}, and for modeling the materials hosting the SkL~\cite{takagi2022square, hayami2023orthorhombic}.

Then, we simplify the model by extracting the dominant $\bm{q}$ component of the interaction, which is given by 
\begin{align}
\label{eq: Ham}
\mathcal{H}=  &-2J \sum_{\nu} (\bm{S}_{\bm{Q}_\nu} \cdot  \bm{S}_{-\bm{Q}_\nu}+I^z S^{z}_{\bm{Q}_\nu}S^{z}_{-\bm{Q}_\nu}) \nonumber \\
&+2\frac{K}{N} \sum_{\nu} (\bm{S}_{\bm{Q}_\nu}\cdot \bm{S}_{-\bm{Q}_\nu}+I^z S^{z}_{\bm{Q}_\nu}S^{z}_{-\bm{Q}_\nu})^2   -  H \sum_{i}   S^z_{i}. 
\end{align}
We regard the interactions at $\bm{Q}_{\nu}$ for $\nu=1$--$4$ as the dominant ones: $\bm{Q}_1=(\pi/4, 3\pi/4)$, $\bm{Q}_2=(3\pi/4, -\pi/4)$, $\bm{Q}_3=(3\pi/4, \pi/4)$, and $\bm{Q}_4=(-\pi/4, 3\pi/4)$, as schematically shown in Fig.~\ref{fig: Qvec}. 
$\bm{Q}_1$--$\bm{Q}_4$ correspond to the octuple-period wave vectors.  
It is noted that $\bm{Q}_{\nu}$ are symmetry-equivalent to each other, for example, $\bm{Q}_{1}$ is connected to $\bm{Q}_2$ by the fourfold rotational symmetry and $\bm{Q}_1$ is connected to $\bm{Q}_3$ by the mirror symmetry around the [110] direction. 
This means that the interaction constants $J_{\bm{q}}$ and $K_{\bm{q}}$ at $\bm{Q}_\nu$ are the same; we set $J \equiv J_{\bm{Q}_\nu}$ and $K \equiv K_{\bm{Q}_\nu}$ and choose $J$ as the energy unit, i.e., $J=1$.  
As detailed in Sec.~\ref{sec: Octuple-period spin density waves}, the choice of octuple-period $\bm{Q}_{\nu}$ is important to induce the instability toward the 4$Q$ CBL state. 
The prefactor 2 in the first and second terms represents the contribution from $-\bm{Q}_{\nu}$.

In the end, the ground-state phase diagram is numerically obtained by minimizing the energy of the model in Eq.~(\ref{eq: Ham}) on the square lattice under the periodic boundary condition, where the system size is taken at $N=16^2$. 
To unbiasedly determine the phase diagram, we adopt simulated annealing based on the standard Metropolis local updates in real space in the following manner. 
First, we start from a random spin configuration at a high temperature $T_0=1.5$. 
By gradually reducing the temperature with a ratio $T_{n+1}=0.999999 T_{n}$ in each Monte Carlo sweep ($T_{n}$ is the $n$th-step temperature), we try to avoid metastable spin configurations. 
Once the temperature reaches the final temperature taken at $T_f=0.0001$, we perform $10^5$-$10^6$ Monte Carlo sweeps for measurements at the final temperature after $10^5$-$10^6$ steps for thermalization.  
In order to determine the phase boundaries between different magnetic states, we also start from the spin patterns obtained at low temperatures by the above procedure; in this case, we start the simulations from $T_0=0.01$. 

To distinguish the multiple-$Q$ state from the single-$Q$ state, we calculate the spin structure factor, which is given by  
\begin{align}
S_s^{\eta\eta}(\bm{q})&= \frac{1}{N} \sum_{i,j} S_i^{\eta} S_j^{\eta}  e^{i \bm{q}\cdot (\bm{r}_i-\bm{r}_j)},
\end{align}
for $\eta=x,y,z$; $\bm{r}_i$ is the position vector at site $i$. 
We also compute the $\bm{Q}_\nu$ component of the magnetic moments as 
\begin{equation}
m^{\eta}_{\bm{Q}_\nu}=\sqrt{\frac{S^{\eta \eta}_s(\bm{Q}_\nu)}
{N}}. 
\end{equation}
For the in-plane spin component, we use the notation 
\begin{align}
m^{xy}_{\bm{Q}_\nu}= \sqrt{\frac{S_s^{xx}(\bm{Q}_\nu)+S_s^{yy}(\bm{Q}_\nu)}{N}}. 
\end{align}
In addition, the net magnetization along the $z$ direction is given by 
\begin{equation}
M^z=
\frac{1}{N}
\sum_i S_i^z. 
\end{equation}

\section{Results}
\label{sec: Results}

In this section, we discuss the stability of the 4$Q$ CBL state. 
We show that the geometry and symmetry of the ordering wave vectors $\bm{Q}_1$--$\bm{Q}_4$ play an important role in inducing the multiple-$Q$ superposition in Sec.~\ref{sec: Octuple-period spin density waves}; the energy for the helical state becomes higher than that for the double-$Q$ collinear (2$Q$ collinear) and 4$Q$ CBL states for $I^z>0$. 
Then, we show that the 4$Q$ CBL state is chosen as the ground state by additionally considering the effect of the positive biquadratic interaction in Sec.~\ref{sec: Instability toward checkerboard bubble lattice}. 
Lastly, we construct the magnetic-field phase diagram while changing the easy-axis two-spin anisotropic interaction ($I^z$) in Sec.~\ref{sec: Phase diagram under external magnetic field}. 

\subsection{Octuple-period spin density waves}
\label{sec: Octuple-period spin density waves}

\begin{figure}[t!]
\begin{center}
\includegraphics[width=1.0 \hsize ]{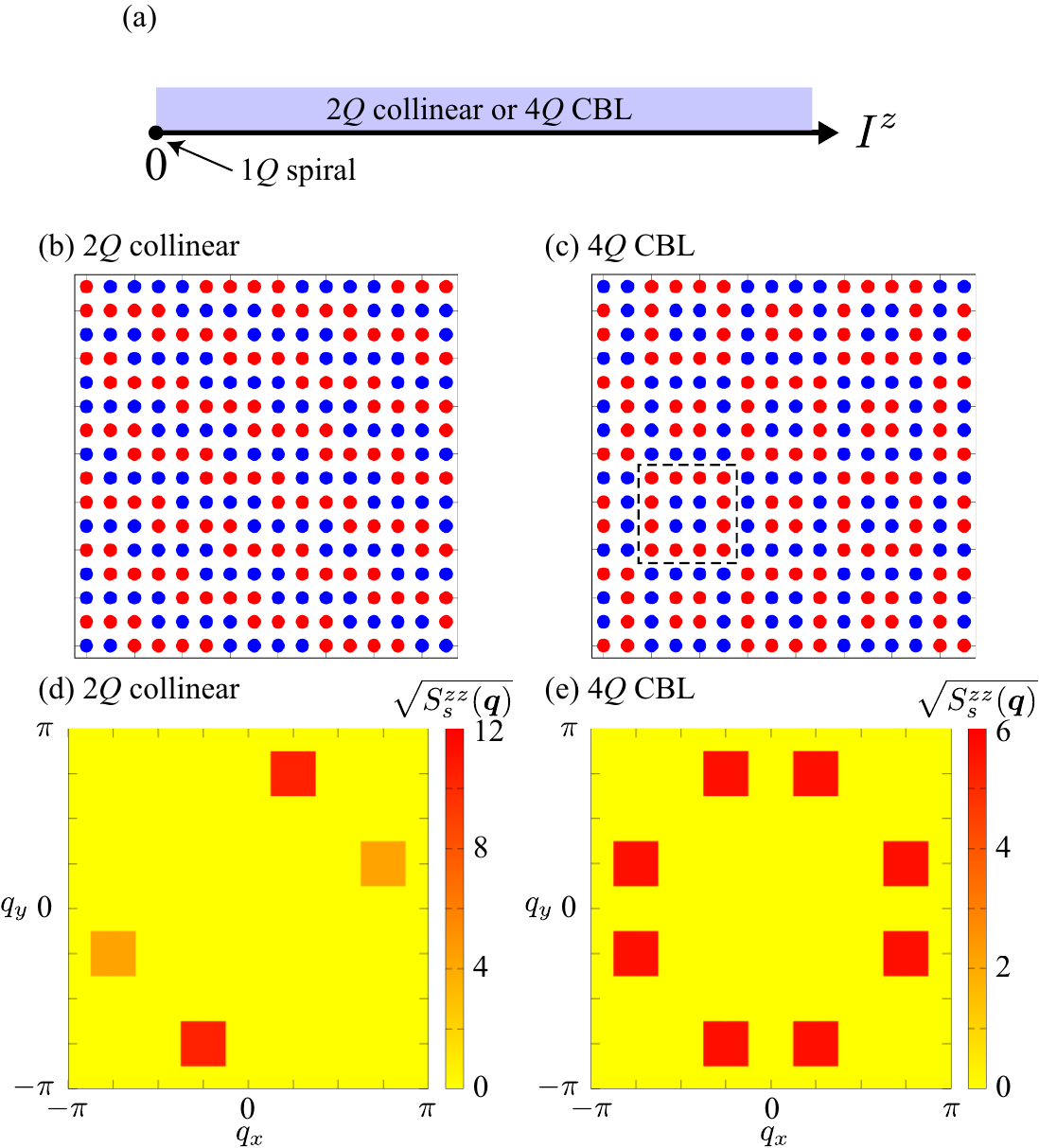} 
\caption{
\label{fig: spin}
(a) Ground-state phase diagram with changing $I^z$ at $K=H=0$. 
The 1$Q$ spiral state only appears at $I^z =0$; it degenerates with the 2$Q$ collinear and 4$Q$ CBL states. 
For $I^z>0$, the 2$Q$ collinear and 4$Q$ CBL states become the ground state. 
Real-space spin configuration of (b) the 2$Q$ collinear state and (c) the 4$Q$ CBL state. 
The red and blue circles represent $S^z_i=+1$ and $S^z_i=-1$, respectively. 
In (c), the dashed square denotes the unit of the bubble. 
The $z$ component of the spin structure factor of (d) the 2$Q$ collinear state and (e) the 4$Q$ CBL state. 
}
\end{center}
\end{figure}

First, let us consider the situation where the effect of the biquadratic interaction is negligible, i.e., $K=0$; the model reduces to the RKKY model with the bilinear exchange interaction.
By further considering the zero-field case $H=0$, the magnetic instability at zero temperature is determined by the ratio of $I^z/J$.  
When $I^z=0$, i.e., the isotropic case, the ground state is given by the single-$Q$ spiral state with $\bm{Q}_\nu$, where the spiral plane is arbitrary. 

Next, we introduce the easy-axis magnetic anisotropy by setting $I^z>0$. 
By performing the simulated annealing in Sec.~\ref{sec: Model and method}, we find that two states with different spin configurations appear against $I^z$, as shown in Fig.~\ref{fig: spin}(a); the 2$Q$ collinear state and the 4$Q$ CBL state. 
It is noted that these two states are energetically degenerate irrespective of the value of $I^z$, although their energy is lower than that in the single-$Q$ spiral state. 
Thus, the multiple-$Q$ instability occurs even without the biquadratic (multiple-spin) interaction or the magnetic field in the presence of an infinitesimally small $I^z$. 

The real-space spin configurations and spin structure factors in the two states are shown in Figs.~\ref{fig: spin}(b)--\ref{fig: spin}(e). 
Both states are characterized by collinear spin textures, as shown in Fig.~\ref{fig: spin}(b) for the 2$Q$ collinear state and Fig.~\ref{fig: spin}(c) for the 4$Q$ CBL state. 
In other words, there are no $xy$-spin components in these two states. 
In the 2$Q$ collinear state, the spin configuration is characterized by a double-$Q$ superposition with different intensities at $\bm{Q}_1$ and $\bm{Q}_3$, as shown in the spin structure factor in Fig.~\ref{fig: spin}(d). 
It is noteworthy that the constituent wave vectors in the 2$Q$ collinear state are connected by the mirror symmetry rather than the rotational symmetry in contrast to conventional multiple-$Q$ states including the SkL. 
Since such a superposition does not occur if the wave vectors are located at high-symmetric lines in the Brillouin zone like $\bm{Q}_\nu \parallel [100] $ or $\bm{Q}_\nu \parallel [110] $~\cite{Yambe_PhysRevB.106.174437}, this result indicates that the low-symmetric ordering wave vectors can give rise to further intriguing multiple-$Q$ states. 
A similar multiple-$Q$ superposition also happens in the 4$Q$ CBL state. 
In this case, a quadruple-$Q$ superposition with equal intensities at $\bm{Q}_1$--$\bm{Q}_4$ appears in the spin structure factor so as to keep fourfold rotational symmetry, as shown in Fig.~\ref{fig: spin}(e). 
Reflecting the fourfold-symmetric spin structure factor, the real-space spin configuration is also fourfold-symmetric, as shown in Fig.~\ref{fig: spin}(c). 
By closely looking at the real-space spin configuration, one finds that the unit of the bubble denoted by the dashed square aligns in a checkerboard way; the ``antibubble" with the opposite $S_i^z$ to the bubble appears next to the bubble so that the total magnetization in the whole system vanishes. 
This is why we call this state the 4$Q$ CBL state. 

We find the analytical expressions of the spin configuration in these two states. 
In the 2$Q$ collinear state, it is given by 
\begin{align}
\label{eq: 2Q}
S_i^z=&\frac{1}{\sqrt{2}} \Big[-\sin \left(\bm{Q}_1 \cdot \bm{r}_i-\frac{\pi}{4} \right)- \sin(\bm{Q}_1 \cdot \bm{r}_i)  \nonumber \\
&+\sin \left(\bm{Q}_3 \cdot \bm{r}_i+\frac{\pi}{4} \right)- \sin(\bm{Q}_3 \cdot \bm{r}_i) \Big],
\end{align}
with $S_i^x=S_i^y=0$. 
Meanwhile, the spin pattern in the 4$Q$ CBL state is given by 
\begin{align}
\label{eq: 4Q}
S_i^z=&\frac{1}{\sqrt{2}} \Big[ \sin \left(\bm{Q}_1 \cdot \bm{r}_i+\frac{\pi}{4} \right)+ \sin(\bm{Q}_2 \cdot \bm{r}_i) \nonumber \\
&-\sin \left(\bm{Q}_3 \cdot \bm{r}_i-\frac{\pi}{4} \right)- \sin(\bm{Q}_4 \cdot \bm{r}_i) \Big],
\end{align}
with $S_i^x=S_i^y=0$. 
It is noted that both spin configurations in Eqs.~(\ref{eq: 2Q}) and (\ref{eq: 4Q}) satisfy $|\bm{S}_i|=1$ at any site without the normalization factor. 

From the expressions in Eqs.~(\ref{eq: 2Q}) and (\ref{eq: 4Q}), one obtains $\bm{Q}_\nu$ component of the magnetic moments. 
In the 2$Q$ collinear state, nonzero $(m^z_{\bm{Q}_\nu})^2$ are given by 
\begin{align}
(m^z_{\bm{Q}_1})^2&= \frac{2+\sqrt{2}}{8}, \nonumber \\
(m^z_{\bm{Q}_3})^2&=\frac{2-\sqrt{2}}{8}, 
\end{align}
and, in the 4$Q$ CBL state, they are given by 
\begin{align}
(m^z_{\bm{Q}_1})^2=(m^z_{\bm{Q}_2})^2=(m^z_{\bm{Q}_3})^2=(m^z_{\bm{Q}_4})^2= \frac{1}{8}. 
\end{align}
From the relation $(m_{-\bm{Q}_\nu})^2=(m_{\bm{Q}_\nu})^2$, both states satisfy $\sum_{\nu}[(m^z_{\bm{Q}_\nu})^2+(m^z_{-\bm{Q}_\nu})^2]=1$, which indicates that there are no contributions from the magnetic moments with the high-harmonic wave vectors. 

\begin{figure}[t!]
\begin{center}
\includegraphics[width=0.8 \hsize ]{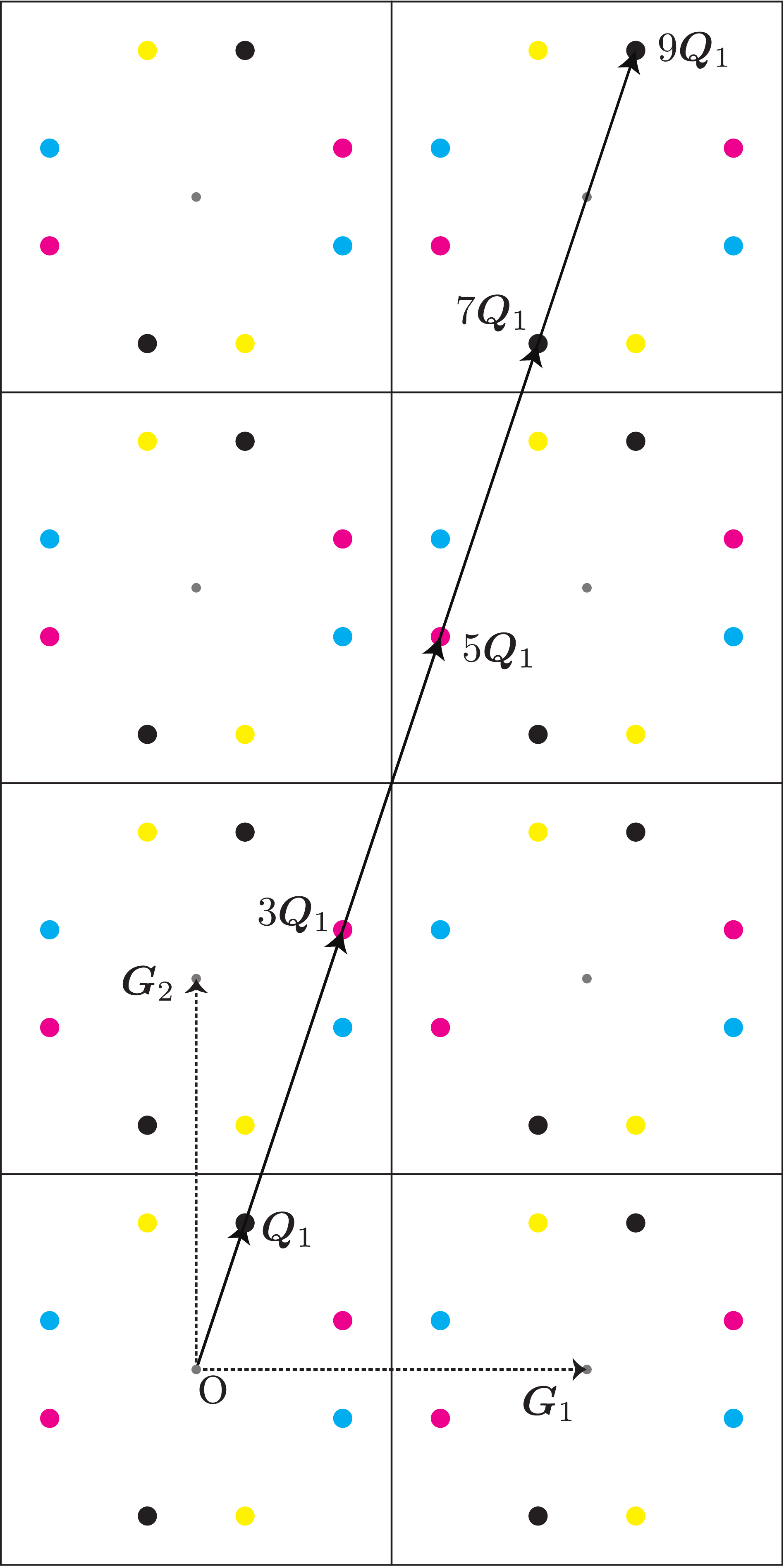} 
\caption{
\label{fig: Qvec2}
Ordering wave vectors at $\bm{Q}_1$, $3\bm{Q}_1$, $5\bm{Q}_1$, $7\bm{Q}_1$, and $9\bm{Q}_1$ in the extended Brillouin zone. 
$\bm{G}_1$ and $\bm{G}_2$ represent the reciprocal lattice vectors. 
}
\end{center}
\end{figure}

It is surprising that no intensities at high-harmonic wave vectors appear in both the 2$Q$ collinear and 4$Q$ CBL states in spite of their multiple-$Q$ structures. 
Although similar situations have been found when the multiple-$Q$ wave vectors lie at the high-symmetric points at the Brillouin zone boundary~\cite{Agterberg_PhysRevB.62.13816, Martin_PhysRevLett.101.156402, Hayami_PhysRevB.90.060402}, the case under the low-symmetric ordering wave vectors is rare; for example, the double-$Q$ SkL consisting of the wave vectors $\bm{q}_1$ and $\bm{q}_2$ with $\bm{q}_1 \perp \bm{q}_2$ exhibits the intensity at $\bm{q}_1+\bm{q}_2$~\cite{Hayami_doi:10.7566/JPSJ.89.103702}. 

The absence of high-harmonic wave-vector contributions is understood from the geometry and symmetry of the constituent ordering wave vectors $\bm{Q}_1$--$\bm{Q}_4$. 
When considering the sinusoidal spiral modulation at the $\bm{Q}_1$ component, the high-harmonic wave vectors contributing to the spin configuration are attributed to $3\bm{Q}_1, 5\bm{Q}_1, 7\bm{Q}_1, \cdots$ owing to the local constraint of the spin length ($|\bm{S}_i|=1$). 
In the present choice of $\bm{Q}_1=(\pi/4, 3\pi/4)$, these high-harmonic wave vectors are reduced to the symmetry-equivalent ones to $\bm{Q}_1$ except for the translation by the reciprocal lattice vectors $\bm{G}_1=(2\pi,0)$ and $\bm{G}_2=(0, 2\pi)$ as follows: 
\begin{align}
3\bm{Q}_1 &=\left( \frac{3\pi}{4}, \frac{9\pi}{4} \right)=\bm{Q}_3 \delta_{\bm{G}_2,\bm{0}}, \nonumber \\
5\bm{Q}_1 &=\left( \frac{5\pi}{4}, \frac{15\pi}{4} \right) = -\bm{Q}_3 \delta_{\bm{G}_1,\bm{0}} \delta_{\bm{G}_2,\bm{0}}, \nonumber \\
7\bm{Q}_1 &=\left( \frac{7\pi}{4}, \frac{21\pi}{4} \right) = -\bm{Q}_1 \delta_{\bm{G}_1,\bm{0}} \delta_{\bm{G}_2,\bm{0}}, \nonumber \\
9\bm{Q}_1 &=\left( \frac{9\pi}{4}, \frac{27\pi}{4} \right) = \bm{Q}_1 \delta_{\bm{G}_1,\bm{0}} \delta_{\bm{G}_2,\bm{0}}, 
\end{align}
where $\delta$ is the Kronecker delta; see also Fig.~\ref{fig: Qvec2}. 
This result indicates that the relation of the octuple-period wave vectors connected by the mirror symmetry is essential, which is characteristic of the low-symmetric wave vectors. 
Similarly, $3\bm{Q}_2, 5\bm{Q}_2, 7\bm{Q}_2, \cdots$ are described by $\pm \bm{Q}_2$ and $\pm \bm{Q}_4$. 
This is why these multiple-$Q$ states have the intensities only at $\bm{Q}_1$, $\bm{Q}_2$, $\bm{Q}_3$, and $\bm{Q}_4$. 
Such a situation does not hold for high-symmetric wave vectors along the $[100]$ and [110] directions and other low-symmetric wave vectors except for octuple-period ones, such as $\bm{q}=(3\pi/4, 3\pi/8)$; the single-$Q$ spiral state is favored instead of multiple-$Q$ states irrespective of the presence/absence of $I^z$. 
In other words, the octuple-period ordering wave vectors play an important role. 

The expressions in Eqs.~(\ref{eq: 2Q}) and (\ref{eq: 4Q}) also mean degenerate energies between the 2$Q$ collinear and 4$Q$ CBL states. 
By calculating the energy from the Hamiltonian in Eq.~(\ref{eq: Ham}), it is given by $J(1+I^z)$ for both states. 
Thus, these states have the same energy within the bilinear RKKY interaction, which indicates that other interactions, such as higher-order interactions, are required to lift their degeneracy, as will be discussed in the next section.

\subsection{Instability toward checkerboard bubble lattice}
\label{sec: Instability toward checkerboard bubble lattice}

\begin{figure}[htb!]
\begin{center}
\includegraphics[width=1.0 \hsize ]{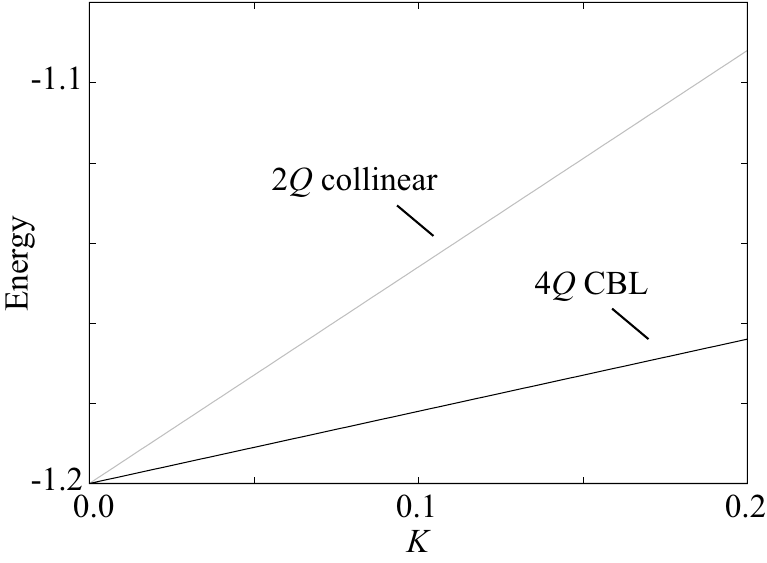} 
\caption{
\label{fig: energy}
$K$ dependence of the energy in the 2$Q$ collinear and 4$Q$ CBL states at $I^z=0.2$ and $H=0$. 
}
\end{center}
\end{figure}

To lift the degeneracy within the RKKY interaction, let us consider the effect of the biquadratic interaction $K$. 
We show the $K$ dependence of the energy for the 2$Q$ collinear and 4$Q$ CBL states at $I^z=0.2$ and $H=0$ in Fig.~\ref{fig: energy}. 
The data clearly show that their degeneracy is lifted for an infinitesimal small $K$; the energy in the 4$Q$ CBL state becomes lower (higher) than that in the 2$Q$ collinear state for $K>0$ ($K<0$). 
This is also understood from the expressions in Eqs.~(\ref{eq: 2Q}) and (\ref{eq: 4Q}): the energy cost by $K$ in the 2$Q$ collinear state is evaluated as $3K(1+I^z)^2/8$, while that in the 4$Q$ CBL state is as $K(1+I^z)^2/8$. 
Thus, nonzero positive (negative) $K$ leads to instability toward the 4$Q$ CBL (2$Q$ collinear) state.

\subsection{Phase diagram under external magnetic field}
\label{sec: Phase diagram under external magnetic field}

\begin{figure}[htb!]
\begin{center}
\includegraphics[width=1.0 \hsize ]{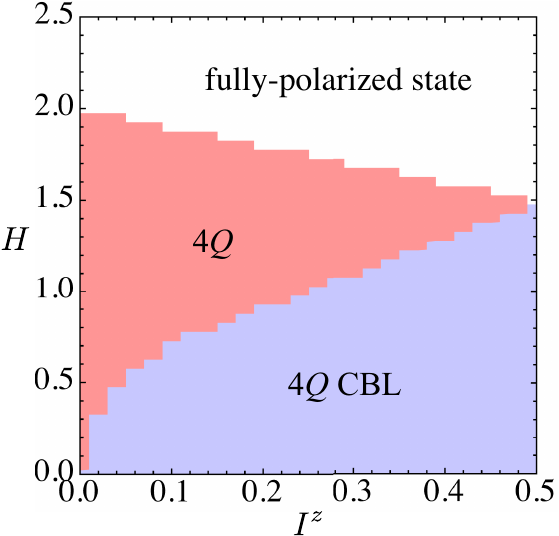} 
\caption{
\label{fig: PD}
Magnetic ground-state phase diagram in the plane of $I^z$ and $H$ at $K=0.05$, which is obtained by the simulated annealing for the model in Eq.~(\ref{eq: Ham}). 
}
\end{center}
\end{figure}

\begin{figure}[htb!]
\begin{center}
\includegraphics[width=1.0 \hsize ]{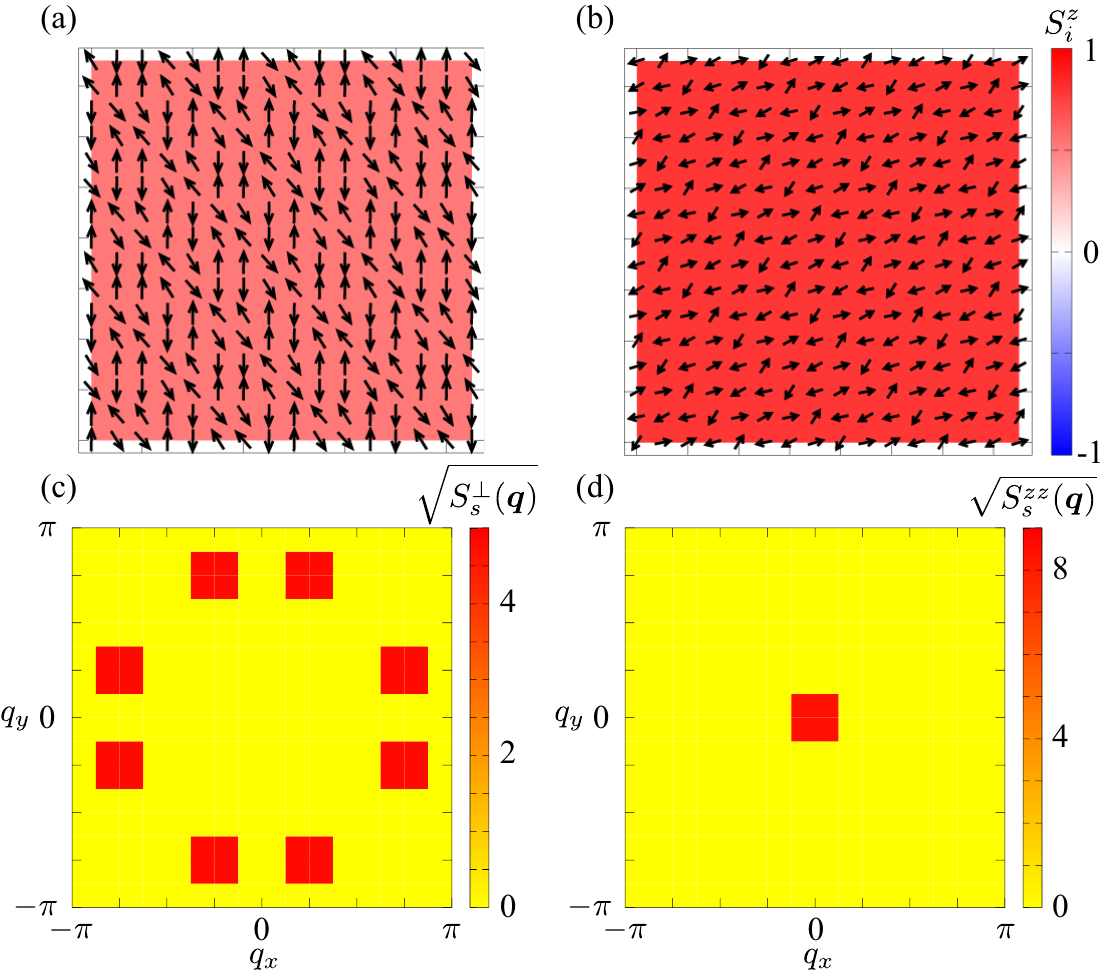} 
\caption{
\label{fig: spin2}
Real-space spin configuration of the $4Q$ state in the high-field region for (a) $H=1$ and (b) $H=1.5$ at $I^z=0.1$. 
The arrows and color denote the $xy$- and $z$-spin components, respectively. 
Square root of the spin structure factor in (c) the $xy$ and (d) $z$ component at $H=1$ and $I^z=0.1$. 
}
\end{center}
\end{figure}

\begin{figure*}[htb!]
\begin{center}
\includegraphics[width=1.0 \hsize ]{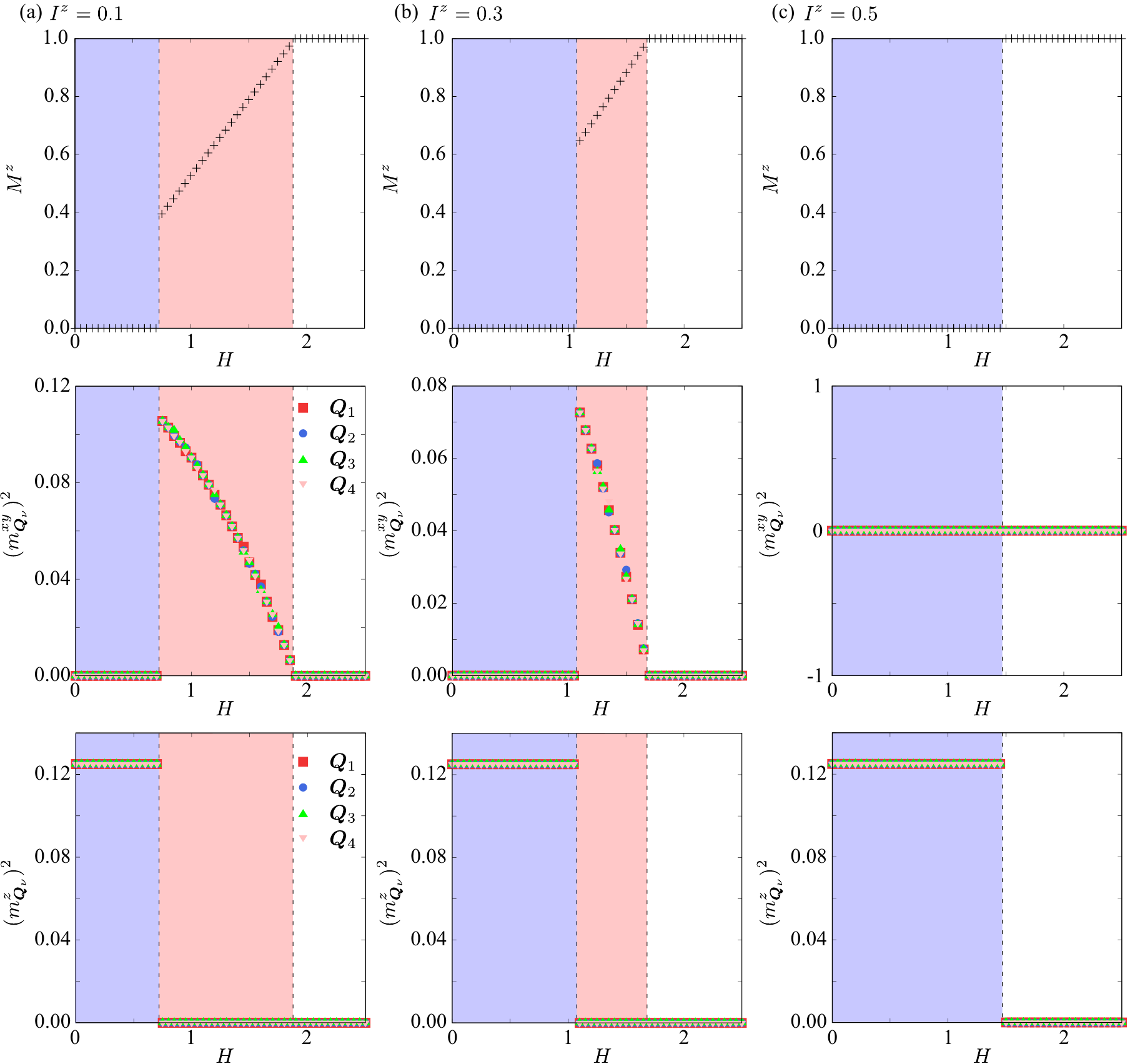} 
\caption{
\label{fig: Mag}
$H$ dependence of the magnetization $M^z$ (top panel), $(m^{xy}_{\bm{Q}_\nu})^2$ (middle panel), and $(m^{z}_{\bm{Q}_\nu})^2$ (bottom panel) for (a) $I^z=0.1$, (b) $I^z=0.3$, and (c) $I^z=0.5$ at $K=0.05$. 
The vertical lines represent the phase boundaries between different magnetic states. 
}
\end{center}
\end{figure*}

We construct the magnetic phase diagram under the external magnetic field at $K=0.05$ in Fig.~\ref{fig: PD}. 
As shown in the phase diagram, the 4$Q$ CBL state is stabilized in the low-field region irrespective of $I^z$, as discussed in Sec.~\ref{sec: Instability toward checkerboard bubble lattice}.  
As $H$ is increased, the 4$Q$ CBL state turns into another quadruple-$Q$ (4$Q$) state, whose real-space spin configuration and the spin structure factor are shown in Fig.~\ref{fig: spin2}. 
We also plot the $H$ dependence of $M^z$, $(m^{xy}_{\bm{Q}_\nu})^2$, and $(m^{z}_{\bm{Q}_\nu})^2$ at several $I^z$ in Fig.~\ref{fig: Mag}.

As shown in the real-space spin configuration in Fig.~\ref{fig: spin2}(a) for $H=1$ and Fig.~\ref{fig: spin2}(b) for $H=1.5$, the 4$Q$ state exhibits a complicated noncoplanar spin texture. 
Similarly to the 4$Q$ CBL state, the 4$Q$ state is characterized by the quadruple-$Q$ peaks with the same intensity in the spin structure factor, although such a peak structure appears in the $xy$-spin component rather than the $z$-spin one, as shown in Figs.~\ref{fig: spin2}(c) and \ref{fig: spin2}(d). 
Thus, this phase transition is regarded as a spin-flop transition, where the spin modulation changes from the $z$ direction parallel to $H$ to the $xy$ plane perpendicular to $H$. 
It is noted that all the intensities in the spin structure factor are located at $\bm{Q}_\nu$ so that the contributions from other wave vectors are absent, which might be attributed to the fact that the high-harmonic wave vectors like $3\bm{Q}_1$, $5\bm{Q}_1$, $7\bm{Q}_1$, and so on, correspond to the symmetry-equivalent ones, as discussed in Sec.~\ref{sec: Octuple-period spin density waves}. 
It is also noted that there is no net scalar spin chirality in this state. 

From the result in Fig.~\ref{fig: Mag}, the transition between the 4$Q$ CBL and $4Q$ states is of first order with jumps of $M^z$, $(m^{xy}_{\bm{Q}_\nu})^2$, and $(m^{z}_{\bm{Q}_\nu})^2$ in Figs.~\ref{fig: Mag}(a) and \ref{fig: Mag}(b). 
Especially, one finds the magnetic plateau in the zero-field region, indicating a finite spin gap owing to the collinear spin configuration along the easy-axis direction. 
By further increasing $H$ in the 4$Q$ state, $(m^{xy}_{\bm{Q}_\nu})^2$ is gradually reduced and this state continuously changes into the fully-polarized state with $S_i^z=1$. 

When $I^z$ is increased, the region of the $4Q$ state is shrunk and vanishes at $I^z = 0.5$; the direct transition from the 4$Q$ CBL state to the fully-polarized state occurs by a drastic jump of the magnetization from $M^z=0$ to $M^z=1$, as shown in Fig.~\ref{fig: Mag}(c). 
The critical value of $H$ between this direct transition is given by $-J(1+I^z)+K(1+I^z)^2/8$.

\section{Discussion}
\label{sec: Discussion}

\subsection{Charge density wave}
\label{sec: Charge density wave}

\begin{figure}[htb!]
\begin{center}
\includegraphics[width=1.0 \hsize ]{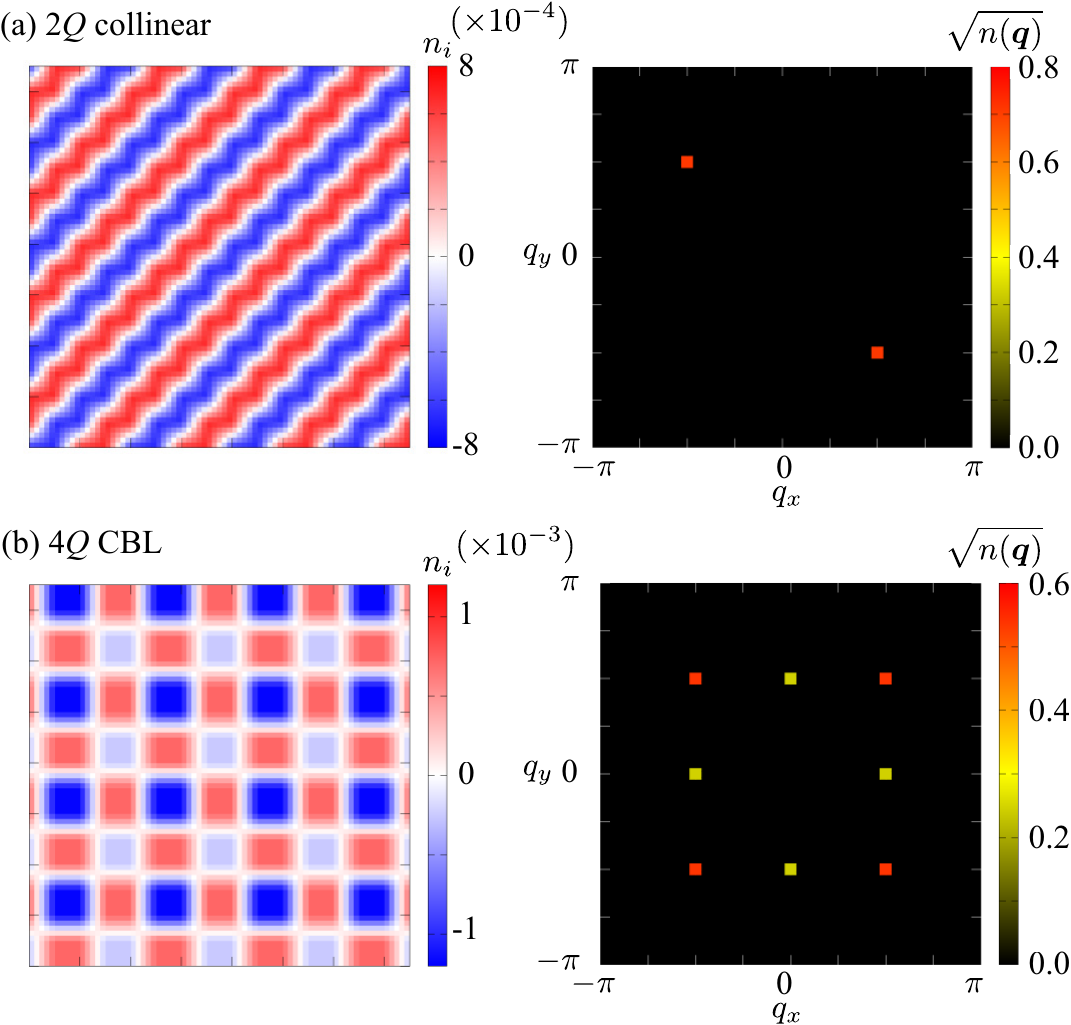} 
\caption{
\label{fig: CDW}
(Left panel) Real-space distributions of the local charge density measured from the average density in (a) the 2$Q$ collinear state and (b) the 4$Q$ CBL state at $J_{\rm K}=0.2$ and $\mu=0.5$. 
(Right panel) The square root of the charge structure factor except for the $\bm{q}=\bm{0}$ component. 
}
\end{center}
\end{figure}

Since the 4$Q$ CBL state exhibits no intensity at high-harmonic wave vectors, such as $\bm{Q}_1+\bm{Q}_2$, it is difficult to distinguish the 4$Q$ CBL state from the multi-domain single-$Q$ state or the multi-domain $2Q$ collinear state in diffraction experiments, such as the small-angle neutron scattering, in contrast to the conventional multiple-$Q$ states like the SkL. 
We here propose how to identify the 4$Q$ CBL state by other measurements. 
One of the powerful measurements is the spectroscopic-imaging scanning tunneling microscopy measurement, which has been recently used to detect multiple-$Q$ states in GdRu$_2$Si$_2$~\cite{Yasui2020imaging}. 
This method enables us to derive the charge density modulations in real and momentum spaces, which are brought about by the formation of magnetic textures~\cite{Hayami_PhysRevB.104.144404}. 

As an example, we show the charge-density distributions in the 2$Q$ 
collinear state and the 4$Q$ CBL state in both real and momentum spaces in Figs.~\ref{fig: CDW}(a) and \ref{fig: CDW}(b), respectively. 
Here, we calculate these quantities for the original Kondo lattice model, which is given by 
\begin{align}
\label{eq: HamKLM}
\mathcal{H} = 
-t \sum_{i, j,  \sigma}  c^{\dagger}_{i\sigma}c_{j \sigma}
+J_{\rm K} \sum_{i} \bm{s}_i \cdot \bm{S}_i,  
\end{align}
where $c^{\dagger}_{i\sigma}$ and $c_{i \sigma}$ are creation and annihilation operators of an itinerant electron at site $i$ and spin $\sigma$. 
For simplicity, we consider the nearest-neighbor hopping $t=1$ in the first term and we set $J_{\rm K}=0.2$; $\bm{s}_i$ is the itinerant electron spin and $\bm{S}_i$ is the localized spins. 
For $\bm{S}_i$, we substitute the expressions in Eqs.~(\ref{eq: 2Q}) or (\ref{eq: 4Q}). 
The local charge density measured from the average density is given by $n_i = \sum_{\sigma}\langle c^{\dagger}_{i\sigma} c_{i \sigma} \rangle- n^{\rm ave} $ ($n^{\rm ave}$ is the average density) and the charge structure factor is given by 
\begin{align}
n(\bm{q})=\frac{1}{N} \sum_{ij} n_i n_j e^{i \bm{q} \cdot (\bm{r}_i - \bm{r}_j)}. 
\end{align}
We set the chemical potential $\mu=0.5$ without loss of generality. 

As shown by the real-space charge distributions in the left panel of Figs.~\ref{fig: CDW}(a) and \ref{fig: CDW}(b), both 2$Q$ collinear and 4$Q$ CBL states accompany charge modulations but their spatial alignments are different from each other; the modulated pattern breaks the fourfold rotational symmetry for the 2$Q$ collinear state, while it keeps the fourfold rotational symmetry for the 4$Q$ CBL state. 
Such a feature is clearly seen in momentum space, as shown in the right panel of Figs.~\ref{fig: CDW}(a) and \ref{fig: CDW}(b). 
The 2$Q$ collinear state exhibits the intensity at $\bm{Q}^{\rm CDW}_{2Q}=(-\pi/2, \pi/2)$, while the 4$Q$ CBL state exhibits the intensities at $\bm{Q}^{\rm CDW}_{4Q}=(-\pi/2, \pi/2)$, $\bm{Q}'^{\rm CDW}_{4Q}=(\pi/2, \pi/2)$, $\bm{Q}''^{\rm CDW}_{4Q}=(\pi/2, 0)$, and $\bm{Q}'''^{\rm CDW}_{4Q}=(0, \pi/2)$. 

The emergent peak positions in the charge structure factor are analytically derived from the expression as~\cite{Hayami_PhysRevB.104.144404} 
\begin{align}
\label{eq:perturbation_nq}
n_{\bm{q}} \propto \sum_{\bm{q}_1,\bm{q}_2}  
(\bm{S}_{\bm{q}_1}\cdot \bm{S}_{\bm{q}_2}) \delta_{\bm{q}_1+\bm{q}_2,\bm{q}+l_1\bm{G}_1+l_2\bm{G}_2},
\end{align}
where $n_{\bm{q}}$ is the Fourier transform of $n_i$ and $l_{1,2}$ are integers. 
By substituting the expression in Eq.~(\ref{eq: 2Q}) into $(\bm{S}_{\bm{q}_1}\cdot \bm{S}_{\bm{q}_2})$ in Eq.~(\ref{eq:perturbation_nq}), one finds that $n_{\bm{Q}_1+\bm{Q}_3}=0$ but $n_{\bm{Q}_1-\bm{Q}_3} \neq 0$ and $n_{2\bm{Q}_3} \neq 0$ in the 2$Q$ collinear state by using the relation $(-\pi/2, \pi/2)=\bm{Q}_1-\bm{Q}_3=2\bm{Q}_3-\bm{G}_1$. 
When the spin pattern of the 2$Q$ collinear state is characterized by a superposition of the $\bm{Q}_2$ and $\bm{Q}_4$ components instead of the $\bm{Q}_1$ and $\bm{Q}_3$ ones, the charge modulation occurs at the $\bm{Q}_2+\bm{Q}_4=(\pi/2, \pi/2)$ component.  
Thus, the multi-domain structure in the 2$Q$ collinear state exhibits the charge modulations at $\bm{q}=(-\pi/2, \pi/2)$ and $(\pi/2, \pi/2)$. 

In a similar manner, the finite intensities at $\bm{Q}^{\rm CDW}_{4Q}$, $\bm{Q}'^{\rm CDW}_{4Q}$, $\bm{Q}''^{\rm CDW}_{4Q}$, and $\bm{Q}'''^{\rm CDW}_{4Q}$ are explained by nonzero $(\bm{S}_{\bm{q}_1}\cdot \bm{S}_{\bm{q}_2})$ for $\bm{q}_1, \bm{q}_2$=$\bm{Q}_1, \bm{Q}_2, \bm{Q}_3, \bm{Q}_4$. 
It is noteworthy that the 4$Q$ CBL state accompanies the charge density waves at $\bm{Q}''^{\rm CDW}_{4Q}$ and $\bm{Q}'''^{\rm CDW}_{4Q}$, which does not appear in the 2$Q$ collinear state. 
The appearance of $\bm{Q}''^{\rm CDW}_{4Q}$ and $\bm{Q}'''^{\rm CDW}_{4Q}$ is attributed to nonzero contributions from a superposition of $\pm \bm{Q}_1$ and $\pm \bm{Q}_4$ ($\pm \bm{Q}_3$ and $\pm \bm{Q}_2$), such as $\bm{Q}_1-\bm{Q}_4=(\pi/2,0)$. 
Thus, the 2$Q$ collinear and 4$Q$ CBL states are distinguishable by detecting the charge density wave even in the presence of the multi-domain structure.

\subsection{Electronic band structure}
\label{sec: Electronic band structure}

\begin{figure}[htb!]
\begin{center}
\includegraphics[width=1.0 \hsize ]{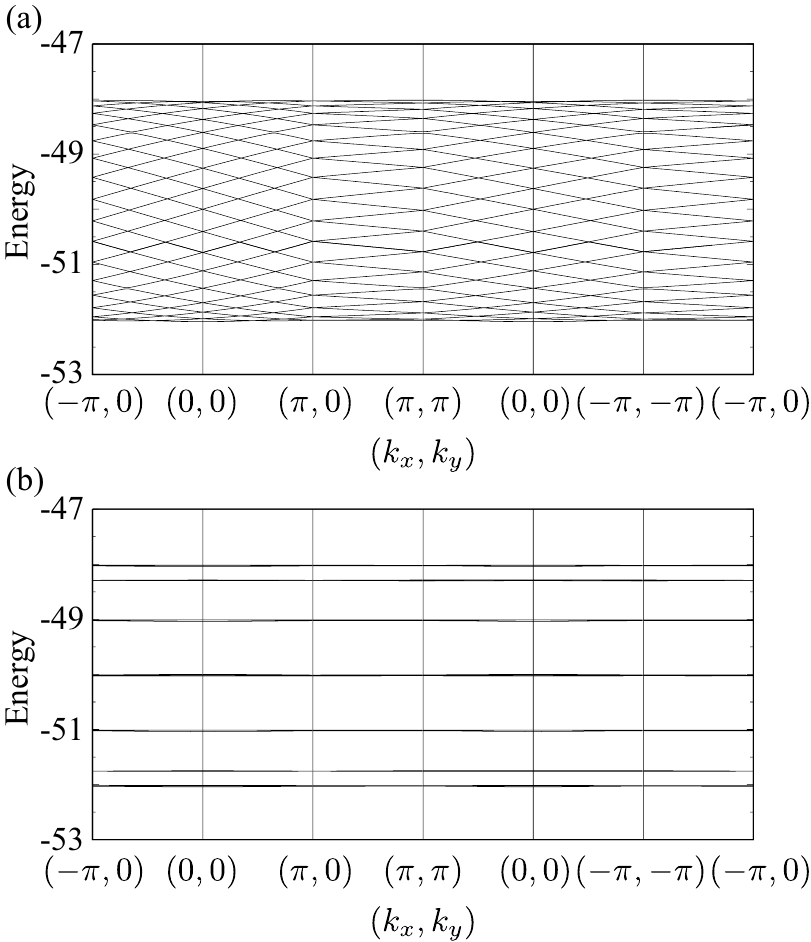} 
\caption{
\label{fig: band}
Electronic band dispersion along the high-symmetric lines under (a) the 2$Q$ collinear state and (b) the 4$Q$ CBL state at $J_{\rm K}=100$. 
}
\end{center}
\end{figure}

Finally, let us discuss a characteristic electronic structure under the 4$Q$ CBL state. 
By closely looking at the real-space spin configuration in Fig.~\ref{fig: spin}(c), the down spins are surrounded by the up spins in the bubble denoted by the dashed square in the 4$Q$ CBL state. 
This indicates the possibility of real-space localization when the itinerant electron spins are strongly coupled to localized spins, i.e., the double-exchange limit~\cite{Zener_PhysRev.82.403, anderson1955considerations}. 
Indeed, one finds the nearly flat band dispersion in the 4$Q$ CBL state, while it does not appear in the 2$Q$ collinear state, as shown in Figs.~\ref{fig: band}(b) and \ref{fig: band}(a), in the strong-coupling regime; we set the lattice constant $a=8$ in this section.
Thus, intriguing transport phenomena can be expected under the  4$Q$ CBL state.

\section{Summary}
\label{sec: Summary}

To summarize, we have investigated multiple-$Q$ bubble-lattice instability consisting of sinusoidal spin density waves on the square lattice. 
By focusing on the geometry and symmetry of the ordering wave vectors in momentum space, we find that octuple-period spin density waves naturally give rise to two types of collinear multiple-$Q$ superpositions, the 2$Q$ collinear and 4$Q$ CBL states, for an infinitesimal easy-axis magnetic anisotropy within the RKKY level. 
In contrast to conventional multiple-$Q$ states consisting of spiral waves like the SkL, no intensities at high-harmonic wave vectors among the constituent ordering wave vectors appear in these states. 
We show that the 4$Q$ CBL state becomes the ground state in the presence of the positive biquadratic interaction by performing numerical and analytical calculations. 
In addition, we show that the 4$Q$ CBL state exhibits the magnetization plateau in the zero-field region owing to the large spin gap. 
Moreover, we demonstrate that the 4$Q$ CBL state accompanies the characteristic charge density waves, which are detected by the spectroscopic-imaging scanning tunneling microscopy measurements. 
We also discuss the nearly flat band structure under the 4$Q$ CBL state in the strong-coupling regime. 
Our results indicate a further possibility of realizing exotic multiple-$Q$ states by controlling the positions of the ordering wave vectors.

\begin{acknowledgments}
This research was supported by JSPS KAKENHI Grants Numbers JP21H01037, JP22H04468, JP22H00101, JP22H01183, JP23K03288, JP23H04869, and by JST PRESTO (JPMJPR20L8). 
Parts of the numerical calculations were performed in the supercomputing systems in ISSP, the University of Tokyo.
\end{acknowledgments}

\appendix

\bibliographystyle{apsrev}
\bibliography{../ref}

\begin{thebibliography}{77}
\expandafter\ifx\csname natexlab\endcsname\relax\def\natexlab#1{#1}\fi
\expandafter\ifx\csname bibnamefont\endcsname\relax
  \def\bibnamefont#1{#1}\fi
\expandafter\ifx\csname bibfnamefont\endcsname\relax
  \def\bibfnamefont#1{#1}\fi
\expandafter\ifx\csname citenamefont\endcsname\relax
  \def\citenamefont#1{#1}\fi
\expandafter\ifx\csname url\endcsname\relax
  \def\url#1{\texttt{#1}}\fi
\expandafter\ifx\csname urlprefix\endcsname\relax\def\urlprefix{URL }\fi
\providecommand{\bibinfo}[2]{#2}
\providecommand{\eprint}[2][]{\url{#2}}

\bibitem[{\citenamefont{R{\"o}{\ss}ler
  et~al.}(2006)\citenamefont{R{\"o}{\ss}ler, Bogdanov, and
  Pfleiderer}}]{rossler2006spontaneous}
\bibinfo{author}{\bibfnamefont{U.~K.} \bibnamefont{R{\"o}{\ss}ler}},
  \bibinfo{author}{\bibfnamefont{A.~N.} \bibnamefont{Bogdanov}},
  \bibnamefont{and}
  \bibinfo{author}{\bibfnamefont{C.}~\bibnamefont{Pfleiderer}},
  \bibinfo{journal}{Nature} \textbf{\bibinfo{volume}{442}},
  \bibinfo{pages}{797} (\bibinfo{year}{2006}).

\bibitem[{\citenamefont{M{\"u}hlbauer et~al.}(2009)\citenamefont{M{\"u}hlbauer,
  Binz, Jonietz, Pfleiderer, Rosch, Neubauer, Georgii, and
  B{\"o}ni}}]{Muhlbauer_2009skyrmion}
\bibinfo{author}{\bibfnamefont{S.}~\bibnamefont{M{\"u}hlbauer}},
  \bibinfo{author}{\bibfnamefont{B.}~\bibnamefont{Binz}},
  \bibinfo{author}{\bibfnamefont{F.}~\bibnamefont{Jonietz}},
  \bibinfo{author}{\bibfnamefont{C.}~\bibnamefont{Pfleiderer}},
  \bibinfo{author}{\bibfnamefont{A.}~\bibnamefont{Rosch}},
  \bibinfo{author}{\bibfnamefont{A.}~\bibnamefont{Neubauer}},
  \bibinfo{author}{\bibfnamefont{R.}~\bibnamefont{Georgii}}, \bibnamefont{and}
  \bibinfo{author}{\bibfnamefont{P.}~\bibnamefont{B{\"o}ni}},
  \bibinfo{journal}{Science} \textbf{\bibinfo{volume}{323}},
  \bibinfo{pages}{915} (\bibinfo{year}{2009}).

\bibitem[{\citenamefont{Yu et~al.}(2010)\citenamefont{Yu, Onose, Kanazawa,
  Park, Han, Matsui, Nagaosa, and Tokura}}]{yu2010real}
\bibinfo{author}{\bibfnamefont{X.~Z.} \bibnamefont{Yu}},
  \bibinfo{author}{\bibfnamefont{Y.}~\bibnamefont{Onose}},
  \bibinfo{author}{\bibfnamefont{N.}~\bibnamefont{Kanazawa}},
  \bibinfo{author}{\bibfnamefont{J.~H.} \bibnamefont{Park}},
  \bibinfo{author}{\bibfnamefont{J.~H.} \bibnamefont{Han}},
  \bibinfo{author}{\bibfnamefont{Y.}~\bibnamefont{Matsui}},
  \bibinfo{author}{\bibfnamefont{N.}~\bibnamefont{Nagaosa}}, \bibnamefont{and}
  \bibinfo{author}{\bibfnamefont{Y.}~\bibnamefont{Tokura}},
  \bibinfo{journal}{Nature} \textbf{\bibinfo{volume}{465}},
  \bibinfo{pages}{901} (\bibinfo{year}{2010}).

\bibitem[{\citenamefont{Yu et~al.}(2011)\citenamefont{Yu, Kanazawa, Onose,
  Kimoto, Zhang, Ishiwata, Matsui, and Tokura}}]{yu2011near}
\bibinfo{author}{\bibfnamefont{X.~Z.} \bibnamefont{Yu}},
  \bibinfo{author}{\bibfnamefont{N.}~\bibnamefont{Kanazawa}},
  \bibinfo{author}{\bibfnamefont{Y.}~\bibnamefont{Onose}},
  \bibinfo{author}{\bibfnamefont{K.}~\bibnamefont{Kimoto}},
  \bibinfo{author}{\bibfnamefont{W.}~\bibnamefont{Zhang}},
  \bibinfo{author}{\bibfnamefont{S.}~\bibnamefont{Ishiwata}},
  \bibinfo{author}{\bibfnamefont{Y.}~\bibnamefont{Matsui}}, \bibnamefont{and}
  \bibinfo{author}{\bibfnamefont{Y.}~\bibnamefont{Tokura}},
  \bibinfo{journal}{Nat. Mater.} \textbf{\bibinfo{volume}{10}},
  \bibinfo{pages}{106} (\bibinfo{year}{2011}).

\bibitem[{\citenamefont{Nagaosa and Tokura}(2013)}]{nagaosa2013topological}
\bibinfo{author}{\bibfnamefont{N.}~\bibnamefont{Nagaosa}} \bibnamefont{and}
  \bibinfo{author}{\bibfnamefont{Y.}~\bibnamefont{Tokura}},
  \bibinfo{journal}{Nat. Nanotechnol.} \textbf{\bibinfo{volume}{8}},
  \bibinfo{pages}{899} (\bibinfo{year}{2013}).

\bibitem[{\citenamefont{Tokura and
  Kanazawa}(2021)}]{Tokura_doi:10.1021/acs.chemrev.0c00297}
\bibinfo{author}{\bibfnamefont{Y.}~\bibnamefont{Tokura}} \bibnamefont{and}
  \bibinfo{author}{\bibfnamefont{N.}~\bibnamefont{Kanazawa}},
  \bibinfo{journal}{Chem. Rev.} \textbf{\bibinfo{volume}{121}},
  \bibinfo{pages}{2857} (\bibinfo{year}{2021}).

\bibitem[{\citenamefont{Yi et~al.}(2009)\citenamefont{Yi, Onoda, Nagaosa, and
  Han}}]{Yi_PhysRevB.80.054416}
\bibinfo{author}{\bibfnamefont{S.~D.} \bibnamefont{Yi}},
  \bibinfo{author}{\bibfnamefont{S.}~\bibnamefont{Onoda}},
  \bibinfo{author}{\bibfnamefont{N.}~\bibnamefont{Nagaosa}}, \bibnamefont{and}
  \bibinfo{author}{\bibfnamefont{J.~H.} \bibnamefont{Han}},
  \bibinfo{journal}{Phys. Rev. B} \textbf{\bibinfo{volume}{80}},
  \bibinfo{pages}{054416} (\bibinfo{year}{2009}).

\bibitem[{\citenamefont{Khanh et~al.}(2020)\citenamefont{Khanh, Nakajima, Yu,
  Gao, Shibata, Hirschberger, Yamasaki, Sagayama, Nakao, Peng
  et~al.}}]{khanh2020nanometric}
\bibinfo{author}{\bibfnamefont{N.~D.} \bibnamefont{Khanh}},
  \bibinfo{author}{\bibfnamefont{T.}~\bibnamefont{Nakajima}},
  \bibinfo{author}{\bibfnamefont{X.}~\bibnamefont{Yu}},
  \bibinfo{author}{\bibfnamefont{S.}~\bibnamefont{Gao}},
  \bibinfo{author}{\bibfnamefont{K.}~\bibnamefont{Shibata}},
  \bibinfo{author}{\bibfnamefont{M.}~\bibnamefont{Hirschberger}},
  \bibinfo{author}{\bibfnamefont{Y.}~\bibnamefont{Yamasaki}},
  \bibinfo{author}{\bibfnamefont{H.}~\bibnamefont{Sagayama}},
  \bibinfo{author}{\bibfnamefont{H.}~\bibnamefont{Nakao}},
  \bibinfo{author}{\bibfnamefont{L.}~\bibnamefont{Peng}}, \bibnamefont{et~al.},
  \bibinfo{journal}{Nat. Nanotechnol.} \textbf{\bibinfo{volume}{15}},
  \bibinfo{pages}{444} (\bibinfo{year}{2020}).

\bibitem[{\citenamefont{Karube et~al.}(2020)\citenamefont{Karube, White,
  Ukleev, Dewhurst, Cubitt, Kikkawa, Tokunaga, R\o{}nnow, Tokura, and
  Taguchi}}]{Karube_PhysRevB.102.064408}
\bibinfo{author}{\bibfnamefont{K.}~\bibnamefont{Karube}},
  \bibinfo{author}{\bibfnamefont{J.~S.} \bibnamefont{White}},
  \bibinfo{author}{\bibfnamefont{V.}~\bibnamefont{Ukleev}},
  \bibinfo{author}{\bibfnamefont{C.~D.} \bibnamefont{Dewhurst}},
  \bibinfo{author}{\bibfnamefont{R.}~\bibnamefont{Cubitt}},
  \bibinfo{author}{\bibfnamefont{A.}~\bibnamefont{Kikkawa}},
  \bibinfo{author}{\bibfnamefont{Y.}~\bibnamefont{Tokunaga}},
  \bibinfo{author}{\bibfnamefont{H.~M.} \bibnamefont{R\o{}nnow}},
  \bibinfo{author}{\bibfnamefont{Y.}~\bibnamefont{Tokura}}, \bibnamefont{and}
  \bibinfo{author}{\bibfnamefont{Y.}~\bibnamefont{Taguchi}},
  \bibinfo{journal}{Phys. Rev. B} \textbf{\bibinfo{volume}{102}},
  \bibinfo{pages}{064408} (\bibinfo{year}{2020}).

\bibitem[{\citenamefont{Khanh et~al.}(2022)\citenamefont{Khanh, Nakajima,
  Hayami, Gao, Yamasaki, Sagayama, Nakao, Takagi, Motome, Tokura
  et~al.}}]{khanh2022zoology}
\bibinfo{author}{\bibfnamefont{N.~D.} \bibnamefont{Khanh}},
  \bibinfo{author}{\bibfnamefont{T.}~\bibnamefont{Nakajima}},
  \bibinfo{author}{\bibfnamefont{S.}~\bibnamefont{Hayami}},
  \bibinfo{author}{\bibfnamefont{S.}~\bibnamefont{Gao}},
  \bibinfo{author}{\bibfnamefont{Y.}~\bibnamefont{Yamasaki}},
  \bibinfo{author}{\bibfnamefont{H.}~\bibnamefont{Sagayama}},
  \bibinfo{author}{\bibfnamefont{H.}~\bibnamefont{Nakao}},
  \bibinfo{author}{\bibfnamefont{R.}~\bibnamefont{Takagi}},
  \bibinfo{author}{\bibfnamefont{Y.}~\bibnamefont{Motome}},
  \bibinfo{author}{\bibfnamefont{Y.}~\bibnamefont{Tokura}},
  \bibnamefont{et~al.}, \bibinfo{journal}{Adv. Sci.}
  \textbf{\bibinfo{volume}{9}}, \bibinfo{pages}{2105452}
  (\bibinfo{year}{2022}).

\bibitem[{\citenamefont{Seki et~al.}(2012)\citenamefont{Seki, Kim, Inosov,
  Georgii, Keimer, Ishiwata, and Tokura}}]{Seki_PhysRevB.85.220406}
\bibinfo{author}{\bibfnamefont{S.}~\bibnamefont{Seki}},
  \bibinfo{author}{\bibfnamefont{J.-H.} \bibnamefont{Kim}},
  \bibinfo{author}{\bibfnamefont{D.~S.} \bibnamefont{Inosov}},
  \bibinfo{author}{\bibfnamefont{R.}~\bibnamefont{Georgii}},
  \bibinfo{author}{\bibfnamefont{B.}~\bibnamefont{Keimer}},
  \bibinfo{author}{\bibfnamefont{S.}~\bibnamefont{Ishiwata}}, \bibnamefont{and}
  \bibinfo{author}{\bibfnamefont{Y.}~\bibnamefont{Tokura}},
  \bibinfo{journal}{Phys. Rev. B} \textbf{\bibinfo{volume}{85}},
  \bibinfo{pages}{220406} (\bibinfo{year}{2012}).

\bibitem[{\citenamefont{Hayami and Yambe}(2023)}]{Hayami_PhysRevB.107.174435}
\bibinfo{author}{\bibfnamefont{S.}~\bibnamefont{Hayami}} \bibnamefont{and}
  \bibinfo{author}{\bibfnamefont{R.}~\bibnamefont{Yambe}},
  \bibinfo{journal}{Phys. Rev. B} \textbf{\bibinfo{volume}{107}},
  \bibinfo{pages}{174435} (\bibinfo{year}{2023}).

\bibitem[{\citenamefont{Ishiwata et~al.}(2011)\citenamefont{Ishiwata, Tokunaga,
  Kaneko, Okuyama, Tokunaga, Wakimoto, Kakurai, Arima, Taguchi, and
  Tokura}}]{Ishiwata_PhysRevB.84.054427}
\bibinfo{author}{\bibfnamefont{S.}~\bibnamefont{Ishiwata}},
  \bibinfo{author}{\bibfnamefont{M.}~\bibnamefont{Tokunaga}},
  \bibinfo{author}{\bibfnamefont{Y.}~\bibnamefont{Kaneko}},
  \bibinfo{author}{\bibfnamefont{D.}~\bibnamefont{Okuyama}},
  \bibinfo{author}{\bibfnamefont{Y.}~\bibnamefont{Tokunaga}},
  \bibinfo{author}{\bibfnamefont{S.}~\bibnamefont{Wakimoto}},
  \bibinfo{author}{\bibfnamefont{K.}~\bibnamefont{Kakurai}},
  \bibinfo{author}{\bibfnamefont{T.}~\bibnamefont{Arima}},
  \bibinfo{author}{\bibfnamefont{Y.}~\bibnamefont{Taguchi}}, \bibnamefont{and}
  \bibinfo{author}{\bibfnamefont{Y.}~\bibnamefont{Tokura}},
  \bibinfo{journal}{Phys. Rev. B} \textbf{\bibinfo{volume}{84}},
  \bibinfo{pages}{054427} (\bibinfo{year}{2011}).

\bibitem[{\citenamefont{Tanigaki et~al.}(2015)\citenamefont{Tanigaki, Shibata,
  Kanazawa, Yu, Onose, Park, Shindo, and Tokura}}]{tanigaki2015real}
\bibinfo{author}{\bibfnamefont{T.}~\bibnamefont{Tanigaki}},
  \bibinfo{author}{\bibfnamefont{K.}~\bibnamefont{Shibata}},
  \bibinfo{author}{\bibfnamefont{N.}~\bibnamefont{Kanazawa}},
  \bibinfo{author}{\bibfnamefont{X.}~\bibnamefont{Yu}},
  \bibinfo{author}{\bibfnamefont{Y.}~\bibnamefont{Onose}},
  \bibinfo{author}{\bibfnamefont{H.~S.} \bibnamefont{Park}},
  \bibinfo{author}{\bibfnamefont{D.}~\bibnamefont{Shindo}}, \bibnamefont{and}
  \bibinfo{author}{\bibfnamefont{Y.}~\bibnamefont{Tokura}},
  \bibinfo{journal}{Nano Lett.} \textbf{\bibinfo{volume}{15}},
  \bibinfo{pages}{5438} (\bibinfo{year}{2015}).

\bibitem[{\citenamefont{Kanazawa et~al.}(2017)\citenamefont{Kanazawa, Seki, and
  Tokura}}]{kanazawa2017noncentrosymmetric}
\bibinfo{author}{\bibfnamefont{N.}~\bibnamefont{Kanazawa}},
  \bibinfo{author}{\bibfnamefont{S.}~\bibnamefont{Seki}}, \bibnamefont{and}
  \bibinfo{author}{\bibfnamefont{Y.}~\bibnamefont{Tokura}},
  \bibinfo{journal}{Adv. Mater.} \textbf{\bibinfo{volume}{29}},
  \bibinfo{pages}{1603227} (\bibinfo{year}{2017}).

\bibitem[{\citenamefont{Fujishiro et~al.}(2019)\citenamefont{Fujishiro,
  Kanazawa, Nakajima, Yu, Ohishi, Kawamura, Kakurai, Arima, Mitamura, Miyake
  et~al.}}]{fujishiro2019topological}
\bibinfo{author}{\bibfnamefont{Y.}~\bibnamefont{Fujishiro}},
  \bibinfo{author}{\bibfnamefont{N.}~\bibnamefont{Kanazawa}},
  \bibinfo{author}{\bibfnamefont{T.}~\bibnamefont{Nakajima}},
  \bibinfo{author}{\bibfnamefont{X.~Z.} \bibnamefont{Yu}},
  \bibinfo{author}{\bibfnamefont{K.}~\bibnamefont{Ohishi}},
  \bibinfo{author}{\bibfnamefont{Y.}~\bibnamefont{Kawamura}},
  \bibinfo{author}{\bibfnamefont{K.}~\bibnamefont{Kakurai}},
  \bibinfo{author}{\bibfnamefont{T.}~\bibnamefont{Arima}},
  \bibinfo{author}{\bibfnamefont{H.}~\bibnamefont{Mitamura}},
  \bibinfo{author}{\bibfnamefont{A.}~\bibnamefont{Miyake}},
  \bibnamefont{et~al.}, \bibinfo{journal}{Nat. Commun.}
  \textbf{\bibinfo{volume}{10}}, \bibinfo{pages}{1059} (\bibinfo{year}{2019}).

\bibitem[{\citenamefont{Ishiwata et~al.}(2020)\citenamefont{Ishiwata, Nakajima,
  Kim, Inosov, Kanazawa, White, Gavilano, Georgii, Seemann, Brandl
  et~al.}}]{Ishiwata_PhysRevB.101.134406}
\bibinfo{author}{\bibfnamefont{S.}~\bibnamefont{Ishiwata}},
  \bibinfo{author}{\bibfnamefont{T.}~\bibnamefont{Nakajima}},
  \bibinfo{author}{\bibfnamefont{J.-H.} \bibnamefont{Kim}},
  \bibinfo{author}{\bibfnamefont{D.~S.} \bibnamefont{Inosov}},
  \bibinfo{author}{\bibfnamefont{N.}~\bibnamefont{Kanazawa}},
  \bibinfo{author}{\bibfnamefont{J.~S.} \bibnamefont{White}},
  \bibinfo{author}{\bibfnamefont{J.~L.} \bibnamefont{Gavilano}},
  \bibinfo{author}{\bibfnamefont{R.}~\bibnamefont{Georgii}},
  \bibinfo{author}{\bibfnamefont{K.~M.} \bibnamefont{Seemann}},
  \bibinfo{author}{\bibfnamefont{G.}~\bibnamefont{Brandl}},
  \bibnamefont{et~al.}, \bibinfo{journal}{Phys. Rev. B}
  \textbf{\bibinfo{volume}{101}}, \bibinfo{pages}{134406}
  (\bibinfo{year}{2020}).

\bibitem[{\citenamefont{Rogge et~al.}(2019)\citenamefont{Rogge, Green, Sutarto,
  and May}}]{Rogge_PhysRevMaterials.3.084404}
\bibinfo{author}{\bibfnamefont{P.~C.} \bibnamefont{Rogge}},
  \bibinfo{author}{\bibfnamefont{R.~J.} \bibnamefont{Green}},
  \bibinfo{author}{\bibfnamefont{R.}~\bibnamefont{Sutarto}}, \bibnamefont{and}
  \bibinfo{author}{\bibfnamefont{S.~J.} \bibnamefont{May}},
  \bibinfo{journal}{Phys. Rev. Materials} \textbf{\bibinfo{volume}{3}},
  \bibinfo{pages}{084404} (\bibinfo{year}{2019}).

\bibitem[{\citenamefont{Kamiya and Batista}(2014)}]{Kamiya_PhysRevX.4.011023}
\bibinfo{author}{\bibfnamefont{Y.}~\bibnamefont{Kamiya}} \bibnamefont{and}
  \bibinfo{author}{\bibfnamefont{C.~D.} \bibnamefont{Batista}},
  \bibinfo{journal}{Phys. Rev. X} \textbf{\bibinfo{volume}{4}},
  \bibinfo{pages}{011023} (\bibinfo{year}{2014}).

\bibitem[{\citenamefont{Liu et~al.}(2016)\citenamefont{Liu, Yu, and
  Wang}}]{Liu_PhysRevB.94.174424}
\bibinfo{author}{\bibfnamefont{C.}~\bibnamefont{Liu}},
  \bibinfo{author}{\bibfnamefont{R.}~\bibnamefont{Yu}}, \bibnamefont{and}
  \bibinfo{author}{\bibfnamefont{X.}~\bibnamefont{Wang}},
  \bibinfo{journal}{Phys. Rev. B} \textbf{\bibinfo{volume}{94}},
  \bibinfo{pages}{174424} (\bibinfo{year}{2016}).

\bibitem[{\citenamefont{Seabra et~al.}(2016)\citenamefont{Seabra, Sindzingre,
  Momoi, and Shannon}}]{Seabra_PhysRevB.93.085132}
\bibinfo{author}{\bibfnamefont{L.}~\bibnamefont{Seabra}},
  \bibinfo{author}{\bibfnamefont{P.}~\bibnamefont{Sindzingre}},
  \bibinfo{author}{\bibfnamefont{T.}~\bibnamefont{Momoi}}, \bibnamefont{and}
  \bibinfo{author}{\bibfnamefont{N.}~\bibnamefont{Shannon}},
  \bibinfo{journal}{Phys. Rev. B} \textbf{\bibinfo{volume}{93}},
  \bibinfo{pages}{085132} (\bibinfo{year}{2016}).

\bibitem[{\citenamefont{Hayami et~al.}(2016{\natexlab{a}})\citenamefont{Hayami,
  Lin, Kamiya, and Batista}}]{Hayami_PhysRevB.94.174420}
\bibinfo{author}{\bibfnamefont{S.}~\bibnamefont{Hayami}},
  \bibinfo{author}{\bibfnamefont{S.-Z.} \bibnamefont{Lin}},
  \bibinfo{author}{\bibfnamefont{Y.}~\bibnamefont{Kamiya}}, \bibnamefont{and}
  \bibinfo{author}{\bibfnamefont{C.~D.} \bibnamefont{Batista}},
  \bibinfo{journal}{Phys. Rev. B} \textbf{\bibinfo{volume}{94}},
  \bibinfo{pages}{174420} (\bibinfo{year}{2016}{\natexlab{a}}).

\bibitem[{\citenamefont{Chern et~al.}(2017)\citenamefont{Chern, Sizyuk, Price,
  and Perkins}}]{Chern_PhysRevB.95.144427}
\bibinfo{author}{\bibfnamefont{G.-W.} \bibnamefont{Chern}},
  \bibinfo{author}{\bibfnamefont{Y.}~\bibnamefont{Sizyuk}},
  \bibinfo{author}{\bibfnamefont{C.}~\bibnamefont{Price}}, \bibnamefont{and}
  \bibinfo{author}{\bibfnamefont{N.~B.} \bibnamefont{Perkins}},
  \bibinfo{journal}{Phys. Rev. B} \textbf{\bibinfo{volume}{95}},
  \bibinfo{pages}{144427} (\bibinfo{year}{2017}).

\bibitem[{\citenamefont{Takagi et~al.}(2018)\citenamefont{Takagi, White,
  Hayami, Arita, Honecker, R{\o}nnow, Tokura, and Seki}}]{takagi2018multiple}
\bibinfo{author}{\bibfnamefont{R.}~\bibnamefont{Takagi}},
  \bibinfo{author}{\bibfnamefont{J.}~\bibnamefont{White}},
  \bibinfo{author}{\bibfnamefont{S.}~\bibnamefont{Hayami}},
  \bibinfo{author}{\bibfnamefont{R.}~\bibnamefont{Arita}},
  \bibinfo{author}{\bibfnamefont{D.}~\bibnamefont{Honecker}},
  \bibinfo{author}{\bibfnamefont{H.}~\bibnamefont{R{\o}nnow}},
  \bibinfo{author}{\bibfnamefont{Y.}~\bibnamefont{Tokura}}, \bibnamefont{and}
  \bibinfo{author}{\bibfnamefont{S.}~\bibnamefont{Seki}},
  \bibinfo{journal}{Sci. Adv.} \textbf{\bibinfo{volume}{4}},
  \bibinfo{pages}{eaau3402} (\bibinfo{year}{2018}).

\bibitem[{\citenamefont{Hayami et~al.}(2021)\citenamefont{Hayami, Okubo, and
  Motome}}]{hayami2021phase}
\bibinfo{author}{\bibfnamefont{S.}~\bibnamefont{Hayami}},
  \bibinfo{author}{\bibfnamefont{T.}~\bibnamefont{Okubo}}, \bibnamefont{and}
  \bibinfo{author}{\bibfnamefont{Y.}~\bibnamefont{Motome}},
  \bibinfo{journal}{Nat. Commun.} \textbf{\bibinfo{volume}{12}},
  \bibinfo{pages}{6927} (\bibinfo{year}{2021}).

\bibitem[{\citenamefont{Kobayashi and
  Hayami}(2022)}]{Kobayashi_PhysRevB.106.L140406}
\bibinfo{author}{\bibfnamefont{K.}~\bibnamefont{Kobayashi}} \bibnamefont{and}
  \bibinfo{author}{\bibfnamefont{S.}~\bibnamefont{Hayami}},
  \bibinfo{journal}{Phys. Rev. B} \textbf{\bibinfo{volume}{106}},
  \bibinfo{pages}{L140406} (\bibinfo{year}{2022}).

\bibitem[{\citenamefont{Zhang et~al.}(2022)\citenamefont{Zhang, Wang, Dong,
  Zhang, Han, and He}}]{Zhang_PhysRevB.105.024411}
\bibinfo{author}{\bibfnamefont{M.}~\bibnamefont{Zhang}},
  \bibinfo{author}{\bibfnamefont{C.}~\bibnamefont{Wang}},
  \bibinfo{author}{\bibfnamefont{S.}~\bibnamefont{Dong}},
  \bibinfo{author}{\bibfnamefont{H.}~\bibnamefont{Zhang}},
  \bibinfo{author}{\bibfnamefont{Y.}~\bibnamefont{Han}}, \bibnamefont{and}
  \bibinfo{author}{\bibfnamefont{L.}~\bibnamefont{He}}, \bibinfo{journal}{Phys.
  Rev. B} \textbf{\bibinfo{volume}{105}}, \bibinfo{pages}{024411}
  (\bibinfo{year}{2022}).

\bibitem[{\citenamefont{Ishitobi and
  Hattori}(2023)}]{Ishitobi_PhysRevB.107.104413}
\bibinfo{author}{\bibfnamefont{T.}~\bibnamefont{Ishitobi}} \bibnamefont{and}
  \bibinfo{author}{\bibfnamefont{K.}~\bibnamefont{Hattori}},
  \bibinfo{journal}{Phys. Rev. B} \textbf{\bibinfo{volume}{107}},
  \bibinfo{pages}{104413} (\bibinfo{year}{2023}).

\bibitem[{\citenamefont{Solenov et~al.}(2012)\citenamefont{Solenov, Mozyrsky,
  and Martin}}]{Solenov_PhysRevLett.108.096403}
\bibinfo{author}{\bibfnamefont{D.}~\bibnamefont{Solenov}},
  \bibinfo{author}{\bibfnamefont{D.}~\bibnamefont{Mozyrsky}}, \bibnamefont{and}
  \bibinfo{author}{\bibfnamefont{I.}~\bibnamefont{Martin}},
  \bibinfo{journal}{Phys. Rev. Lett.} \textbf{\bibinfo{volume}{108}},
  \bibinfo{pages}{096403} (\bibinfo{year}{2012}).

\bibitem[{\citenamefont{Ozawa et~al.}(2016)\citenamefont{Ozawa, Hayami, Barros,
  Chern, Motome, and Batista}}]{Ozawa_doi:10.7566/JPSJ.85.103703}
\bibinfo{author}{\bibfnamefont{R.}~\bibnamefont{Ozawa}},
  \bibinfo{author}{\bibfnamefont{S.}~\bibnamefont{Hayami}},
  \bibinfo{author}{\bibfnamefont{K.}~\bibnamefont{Barros}},
  \bibinfo{author}{\bibfnamefont{G.-W.} \bibnamefont{Chern}},
  \bibinfo{author}{\bibfnamefont{Y.}~\bibnamefont{Motome}}, \bibnamefont{and}
  \bibinfo{author}{\bibfnamefont{C.~D.} \bibnamefont{Batista}},
  \bibinfo{journal}{J. Phys. Soc. Jpn.} \textbf{\bibinfo{volume}{85}},
  \bibinfo{pages}{103703} (\bibinfo{year}{2016}).

\bibitem[{\citenamefont{Wood et~al.}(2023)\citenamefont{Wood, Khalyavin, Mayoh,
  Bouaziz, Hall, Holt, Orlandi, Manuel, Bl\"ugel, Staunton
  et~al.}}]{Wood_PhysRevB.107.L180402}
\bibinfo{author}{\bibfnamefont{G.~D.~A.} \bibnamefont{Wood}},
  \bibinfo{author}{\bibfnamefont{D.~D.} \bibnamefont{Khalyavin}},
  \bibinfo{author}{\bibfnamefont{D.~A.} \bibnamefont{Mayoh}},
  \bibinfo{author}{\bibfnamefont{J.}~\bibnamefont{Bouaziz}},
  \bibinfo{author}{\bibfnamefont{A.~E.} \bibnamefont{Hall}},
  \bibinfo{author}{\bibfnamefont{S.~J.~R.} \bibnamefont{Holt}},
  \bibinfo{author}{\bibfnamefont{F.}~\bibnamefont{Orlandi}},
  \bibinfo{author}{\bibfnamefont{P.}~\bibnamefont{Manuel}},
  \bibinfo{author}{\bibfnamefont{S.}~\bibnamefont{Bl\"ugel}},
  \bibinfo{author}{\bibfnamefont{J.~B.} \bibnamefont{Staunton}},
  \bibnamefont{et~al.}, \bibinfo{journal}{Phys. Rev. B}
  \textbf{\bibinfo{volume}{107}}, \bibinfo{pages}{L180402}
  (\bibinfo{year}{2023}).

\bibitem[{\citenamefont{Zager et~al.}(2023)\citenamefont{Zager, Fan, Steadman,
  and Plumb}}]{zager2023double}
\bibinfo{author}{\bibfnamefont{B.}~\bibnamefont{Zager}},
  \bibinfo{author}{\bibfnamefont{R.}~\bibnamefont{Fan}},
  \bibinfo{author}{\bibfnamefont{P.}~\bibnamefont{Steadman}}, \bibnamefont{and}
  \bibinfo{author}{\bibfnamefont{K.}~\bibnamefont{Plumb}},
  \bibinfo{journal}{arXiv:2307.03776}  (\bibinfo{year}{2023}).

\bibitem[{\citenamefont{Shimokawa and
  Kawamura}(2019)}]{Shimokawa_PhysRevLett.123.057202}
\bibinfo{author}{\bibfnamefont{T.}~\bibnamefont{Shimokawa}} \bibnamefont{and}
  \bibinfo{author}{\bibfnamefont{H.}~\bibnamefont{Kawamura}},
  \bibinfo{journal}{Phys. Rev. Lett.} \textbf{\bibinfo{volume}{123}},
  \bibinfo{pages}{057202} (\bibinfo{year}{2019}).

\bibitem[{\citenamefont{Lin et~al.}(1973)\citenamefont{Lin, Grundy, and
  Giess}}]{lin1973bubble}
\bibinfo{author}{\bibfnamefont{Y.}~\bibnamefont{Lin}},
  \bibinfo{author}{\bibfnamefont{P.}~\bibnamefont{Grundy}}, \bibnamefont{and}
  \bibinfo{author}{\bibfnamefont{E.}~\bibnamefont{Giess}},
  \bibinfo{journal}{Appl. Phys. Lett.} \textbf{\bibinfo{volume}{23}},
  \bibinfo{pages}{485} (\bibinfo{year}{1973}).

\bibitem[{\citenamefont{Garel and Doniach}(1982)}]{Garel_PhysRevB.26.325}
\bibinfo{author}{\bibfnamefont{T.}~\bibnamefont{Garel}} \bibnamefont{and}
  \bibinfo{author}{\bibfnamefont{S.}~\bibnamefont{Doniach}},
  \bibinfo{journal}{Phys. Rev. B} \textbf{\bibinfo{volume}{26}},
  \bibinfo{pages}{325} (\bibinfo{year}{1982}).

\bibitem[{\citenamefont{Takao}(1983)}]{takao1983study}
\bibinfo{author}{\bibfnamefont{S.}~\bibnamefont{Takao}}, \bibinfo{journal}{J.
  Magn. Magn. Mater.} \textbf{\bibinfo{volume}{31}}, \bibinfo{pages}{1009}
  (\bibinfo{year}{1983}).

\bibitem[{\citenamefont{Jungwirth et~al.}(2016)\citenamefont{Jungwirth, Marti,
  Wadley, and Wunderlich}}]{jungwirth2016antiferromagnetic}
\bibinfo{author}{\bibfnamefont{T.}~\bibnamefont{Jungwirth}},
  \bibinfo{author}{\bibfnamefont{X.}~\bibnamefont{Marti}},
  \bibinfo{author}{\bibfnamefont{P.}~\bibnamefont{Wadley}}, \bibnamefont{and}
  \bibinfo{author}{\bibfnamefont{J.}~\bibnamefont{Wunderlich}},
  \bibinfo{journal}{Nat. Nanotechnol.} \textbf{\bibinfo{volume}{11}},
  \bibinfo{pages}{231} (\bibinfo{year}{2016}).

\bibitem[{\citenamefont{Caretta et~al.}(2018)\citenamefont{Caretta, Mann,
  B{\"u}ttner, Ueda, Pfau, G{\"u}nther, Hessing, Churikova, Klose, Schneider
  et~al.}}]{caretta2018fast}
\bibinfo{author}{\bibfnamefont{L.}~\bibnamefont{Caretta}},
  \bibinfo{author}{\bibfnamefont{M.}~\bibnamefont{Mann}},
  \bibinfo{author}{\bibfnamefont{F.}~\bibnamefont{B{\"u}ttner}},
  \bibinfo{author}{\bibfnamefont{K.}~\bibnamefont{Ueda}},
  \bibinfo{author}{\bibfnamefont{B.}~\bibnamefont{Pfau}},
  \bibinfo{author}{\bibfnamefont{C.~M.} \bibnamefont{G{\"u}nther}},
  \bibinfo{author}{\bibfnamefont{P.}~\bibnamefont{Hessing}},
  \bibinfo{author}{\bibfnamefont{A.}~\bibnamefont{Churikova}},
  \bibinfo{author}{\bibfnamefont{C.}~\bibnamefont{Klose}},
  \bibinfo{author}{\bibfnamefont{M.}~\bibnamefont{Schneider}},
  \bibnamefont{et~al.}, \bibinfo{journal}{Nat. Nanotechnol.}
  \textbf{\bibinfo{volume}{13}}, \bibinfo{pages}{1154} (\bibinfo{year}{2018}).

\bibitem[{\citenamefont{Callen and Josephs}(1971)}]{callen1971dynamics}
\bibinfo{author}{\bibfnamefont{H.}~\bibnamefont{Callen}} \bibnamefont{and}
  \bibinfo{author}{\bibfnamefont{R.~M.} \bibnamefont{Josephs}},
  \bibinfo{journal}{J. Appl. Phys.} \textbf{\bibinfo{volume}{42}},
  \bibinfo{pages}{1977} (\bibinfo{year}{1971}).

\bibitem[{\citenamefont{De~Leeuw et~al.}(1980)\citenamefont{De~Leeuw, Van
  Den~Doel, and Enz}}]{de1980dynamic}
\bibinfo{author}{\bibfnamefont{F.}~\bibnamefont{De~Leeuw}},
  \bibinfo{author}{\bibfnamefont{R.}~\bibnamefont{Van Den~Doel}},
  \bibnamefont{and} \bibinfo{author}{\bibfnamefont{U.}~\bibnamefont{Enz}},
  \bibinfo{journal}{Rep. Prog. Phys.} \textbf{\bibinfo{volume}{43}},
  \bibinfo{pages}{689} (\bibinfo{year}{1980}).

\bibitem[{\citenamefont{Moutafis et~al.}(2009)\citenamefont{Moutafis, Komineas,
  and Bland}}]{Moutafis_PhysRevB.79.224429}
\bibinfo{author}{\bibfnamefont{C.}~\bibnamefont{Moutafis}},
  \bibinfo{author}{\bibfnamefont{S.}~\bibnamefont{Komineas}}, \bibnamefont{and}
  \bibinfo{author}{\bibfnamefont{J.~A.~C.} \bibnamefont{Bland}},
  \bibinfo{journal}{Phys. Rev. B} \textbf{\bibinfo{volume}{79}},
  \bibinfo{pages}{224429} (\bibinfo{year}{2009}).

\bibitem[{\citenamefont{Xiao et~al.}(2020)\citenamefont{Xiao, Morvan, He, Wang,
  Luo, Jiao, Xia, Zhao, and Liu}}]{xiao2020spin}
\bibinfo{author}{\bibfnamefont{Y.}~\bibnamefont{Xiao}},
  \bibinfo{author}{\bibfnamefont{F.}~\bibnamefont{Morvan}},
  \bibinfo{author}{\bibfnamefont{A.}~\bibnamefont{He}},
  \bibinfo{author}{\bibfnamefont{M.}~\bibnamefont{Wang}},
  \bibinfo{author}{\bibfnamefont{H.}~\bibnamefont{Luo}},
  \bibinfo{author}{\bibfnamefont{R.}~\bibnamefont{Jiao}},
  \bibinfo{author}{\bibfnamefont{W.}~\bibnamefont{Xia}},
  \bibinfo{author}{\bibfnamefont{G.}~\bibnamefont{Zhao}}, \bibnamefont{and}
  \bibinfo{author}{\bibfnamefont{J.}~\bibnamefont{Liu}},
  \bibinfo{journal}{Appl. Phys. Lett.} \textbf{\bibinfo{volume}{117}},
  \bibinfo{pages}{132402} (\bibinfo{year}{2020}).

\bibitem[{\citenamefont{Gutzeit et~al.}(2022)\citenamefont{Gutzeit, Kubetzka,
  Haldar, Pralow, Goerzen, Wiesendanger, Heinze, and von
  Bergmann}}]{gutzeit2022nano}
\bibinfo{author}{\bibfnamefont{M.}~\bibnamefont{Gutzeit}},
  \bibinfo{author}{\bibfnamefont{A.}~\bibnamefont{Kubetzka}},
  \bibinfo{author}{\bibfnamefont{S.}~\bibnamefont{Haldar}},
  \bibinfo{author}{\bibfnamefont{H.}~\bibnamefont{Pralow}},
  \bibinfo{author}{\bibfnamefont{M.~A.} \bibnamefont{Goerzen}},
  \bibinfo{author}{\bibfnamefont{R.}~\bibnamefont{Wiesendanger}},
  \bibinfo{author}{\bibfnamefont{S.}~\bibnamefont{Heinze}}, \bibnamefont{and}
  \bibinfo{author}{\bibfnamefont{K.}~\bibnamefont{von Bergmann}},
  \bibinfo{journal}{Nat. Commun.} \textbf{\bibinfo{volume}{13}},
  \bibinfo{pages}{5764} (\bibinfo{year}{2022}).

\bibitem[{\citenamefont{Marcus et~al.}(2018)\citenamefont{Marcus, Kim,
  Tutmaher, Rodriguez-Rivera, Birk, Niedermeyer, Lee, Fisk, and
  Broholm}}]{Marcus_PhysRevLett.120.097201}
\bibinfo{author}{\bibfnamefont{G.~G.} \bibnamefont{Marcus}},
  \bibinfo{author}{\bibfnamefont{D.-J.} \bibnamefont{Kim}},
  \bibinfo{author}{\bibfnamefont{J.~A.} \bibnamefont{Tutmaher}},
  \bibinfo{author}{\bibfnamefont{J.~A.} \bibnamefont{Rodriguez-Rivera}},
  \bibinfo{author}{\bibfnamefont{J.~O.} \bibnamefont{Birk}},
  \bibinfo{author}{\bibfnamefont{C.}~\bibnamefont{Niedermeyer}},
  \bibinfo{author}{\bibfnamefont{H.}~\bibnamefont{Lee}},
  \bibinfo{author}{\bibfnamefont{Z.}~\bibnamefont{Fisk}}, \bibnamefont{and}
  \bibinfo{author}{\bibfnamefont{C.~L.} \bibnamefont{Broholm}},
  \bibinfo{journal}{Phys. Rev. Lett.} \textbf{\bibinfo{volume}{120}},
  \bibinfo{pages}{097201} (\bibinfo{year}{2018}).

\bibitem[{\citenamefont{Park et~al.}(2018)\citenamefont{Park, Sakai, Mackenzie,
  and Hicks}}]{Park_PhysRevB.98.024426}
\bibinfo{author}{\bibfnamefont{J.}~\bibnamefont{Park}},
  \bibinfo{author}{\bibfnamefont{H.}~\bibnamefont{Sakai}},
  \bibinfo{author}{\bibfnamefont{A.~P.} \bibnamefont{Mackenzie}},
  \bibnamefont{and} \bibinfo{author}{\bibfnamefont{C.~W.} \bibnamefont{Hicks}},
  \bibinfo{journal}{Phys. Rev. B} \textbf{\bibinfo{volume}{98}},
  \bibinfo{pages}{024426} (\bibinfo{year}{2018}).

\bibitem[{\citenamefont{Seo et~al.}(2020)\citenamefont{Seo, Wang, Thomas, Rahn,
  Carmo, Ronning, Bauer, dos Reis, Janoschek, Thompson
  et~al.}}]{Seo_PhysRevX.10.011035}
\bibinfo{author}{\bibfnamefont{S.}~\bibnamefont{Seo}},
  \bibinfo{author}{\bibfnamefont{X.}~\bibnamefont{Wang}},
  \bibinfo{author}{\bibfnamefont{S.~M.} \bibnamefont{Thomas}},
  \bibinfo{author}{\bibfnamefont{M.~C.} \bibnamefont{Rahn}},
  \bibinfo{author}{\bibfnamefont{D.}~\bibnamefont{Carmo}},
  \bibinfo{author}{\bibfnamefont{F.}~\bibnamefont{Ronning}},
  \bibinfo{author}{\bibfnamefont{E.~D.} \bibnamefont{Bauer}},
  \bibinfo{author}{\bibfnamefont{R.~D.} \bibnamefont{dos Reis}},
  \bibinfo{author}{\bibfnamefont{M.}~\bibnamefont{Janoschek}},
  \bibinfo{author}{\bibfnamefont{J.~D.} \bibnamefont{Thompson}},
  \bibnamefont{et~al.}, \bibinfo{journal}{Phys. Rev. X}
  \textbf{\bibinfo{volume}{10}}, \bibinfo{pages}{011035}
  (\bibinfo{year}{2020}).

\bibitem[{\citenamefont{Seo et~al.}(2021)\citenamefont{Seo, Hayami, Su, Thomas,
  Ronning, Bauer, Thompson, Lin, and Rosa}}]{seo2021spin}
\bibinfo{author}{\bibfnamefont{S.}~\bibnamefont{Seo}},
  \bibinfo{author}{\bibfnamefont{S.}~\bibnamefont{Hayami}},
  \bibinfo{author}{\bibfnamefont{Y.}~\bibnamefont{Su}},
  \bibinfo{author}{\bibfnamefont{S.~M.} \bibnamefont{Thomas}},
  \bibinfo{author}{\bibfnamefont{F.}~\bibnamefont{Ronning}},
  \bibinfo{author}{\bibfnamefont{E.~D.} \bibnamefont{Bauer}},
  \bibinfo{author}{\bibfnamefont{J.~D.} \bibnamefont{Thompson}},
  \bibinfo{author}{\bibfnamefont{S.-Z.} \bibnamefont{Lin}}, \bibnamefont{and}
  \bibinfo{author}{\bibfnamefont{P.~F.} \bibnamefont{Rosa}},
  \bibinfo{journal}{Commun. Phys.} \textbf{\bibinfo{volume}{4}},
  \bibinfo{pages}{58} (\bibinfo{year}{2021}).

\bibitem[{\citenamefont{Hayami et~al.}(2016{\natexlab{b}})\citenamefont{Hayami,
  Lin, and Batista}}]{Hayami_PhysRevB.93.184413}
\bibinfo{author}{\bibfnamefont{S.}~\bibnamefont{Hayami}},
  \bibinfo{author}{\bibfnamefont{S.-Z.} \bibnamefont{Lin}}, \bibnamefont{and}
  \bibinfo{author}{\bibfnamefont{C.~D.} \bibnamefont{Batista}},
  \bibinfo{journal}{Phys. Rev. B} \textbf{\bibinfo{volume}{93}},
  \bibinfo{pages}{184413} (\bibinfo{year}{2016}{\natexlab{b}}).

\bibitem[{\citenamefont{Hayami}(2020)}]{hayami2020multiple}
\bibinfo{author}{\bibfnamefont{S.}~\bibnamefont{Hayami}}, \bibinfo{journal}{J.
  Magn. Magn. Mater.} \textbf{\bibinfo{volume}{513}}, \bibinfo{pages}{167181}
  (\bibinfo{year}{2020}).

\bibitem[{\citenamefont{Hayami and
  Kato}(2023{\natexlab{a}})}]{hayami2023magnetic}
\bibinfo{author}{\bibfnamefont{S.}~\bibnamefont{Hayami}} \bibnamefont{and}
  \bibinfo{author}{\bibfnamefont{Y.}~\bibnamefont{Kato}},
  \bibinfo{journal}{arXiv:2305.08333}  (\bibinfo{year}{2023}{\natexlab{a}}).

\bibitem[{\citenamefont{Hayami}(2021)}]{Hayami_10.1088/1367-2630/ac3683}
\bibinfo{author}{\bibfnamefont{S.}~\bibnamefont{Hayami}}, \bibinfo{journal}{New
  J. Phys.} \textbf{\bibinfo{volume}{23}}, \bibinfo{pages}{113032}
  (\bibinfo{year}{2021}).

\bibitem[{\citenamefont{Yasui et~al.}(2020)\citenamefont{Yasui, Butler, Khanh,
  Hayami, Nomoto, Hanaguri, Motome, Arita, h.~Arima, Tokura
  et~al.}}]{Yasui2020imaging}
\bibinfo{author}{\bibfnamefont{Y.}~\bibnamefont{Yasui}},
  \bibinfo{author}{\bibfnamefont{C.~J.} \bibnamefont{Butler}},
  \bibinfo{author}{\bibfnamefont{N.~D.} \bibnamefont{Khanh}},
  \bibinfo{author}{\bibfnamefont{S.}~\bibnamefont{Hayami}},
  \bibinfo{author}{\bibfnamefont{T.}~\bibnamefont{Nomoto}},
  \bibinfo{author}{\bibfnamefont{T.}~\bibnamefont{Hanaguri}},
  \bibinfo{author}{\bibfnamefont{Y.}~\bibnamefont{Motome}},
  \bibinfo{author}{\bibfnamefont{R.}~\bibnamefont{Arita}},
  \bibinfo{author}{\bibfnamefont{T.}~\bibnamefont{h.~Arima}},
  \bibinfo{author}{\bibfnamefont{Y.}~\bibnamefont{Tokura}},
  \bibnamefont{et~al.}, \bibinfo{journal}{Nat. Commun.}
  \textbf{\bibinfo{volume}{11}}, \bibinfo{pages}{5925} (\bibinfo{year}{2020}).

\bibitem[{\citenamefont{Hayami and
  Motome}(2021{\natexlab{a}})}]{hayami2021topological}
\bibinfo{author}{\bibfnamefont{S.}~\bibnamefont{Hayami}} \bibnamefont{and}
  \bibinfo{author}{\bibfnamefont{Y.}~\bibnamefont{Motome}},
  \bibinfo{journal}{J. Phys.: Condens. Matter} \textbf{\bibinfo{volume}{33}},
  \bibinfo{pages}{443001} (\bibinfo{year}{2021}{\natexlab{a}}).

\bibitem[{\citenamefont{Hayami et~al.}(2017)\citenamefont{Hayami, Ozawa, and
  Motome}}]{Hayami_PhysRevB.95.224424}
\bibinfo{author}{\bibfnamefont{S.}~\bibnamefont{Hayami}},
  \bibinfo{author}{\bibfnamefont{R.}~\bibnamefont{Ozawa}}, \bibnamefont{and}
  \bibinfo{author}{\bibfnamefont{Y.}~\bibnamefont{Motome}},
  \bibinfo{journal}{Phys. Rev. B} \textbf{\bibinfo{volume}{95}},
  \bibinfo{pages}{224424} (\bibinfo{year}{2017}).

\bibitem[{\citenamefont{Ruderman and Kittel}(1954)}]{Ruderman}
\bibinfo{author}{\bibfnamefont{M.~A.} \bibnamefont{Ruderman}} \bibnamefont{and}
  \bibinfo{author}{\bibfnamefont{C.}~\bibnamefont{Kittel}},
  \bibinfo{journal}{Phys. Rev.} \textbf{\bibinfo{volume}{96}},
  \bibinfo{pages}{99} (\bibinfo{year}{1954}).

\bibitem[{\citenamefont{Kasuya}(1956)}]{Kasuya}
\bibinfo{author}{\bibfnamefont{T.}~\bibnamefont{Kasuya}},
  \bibinfo{journal}{Prog. Theor. Phys.} \textbf{\bibinfo{volume}{16}},
  \bibinfo{pages}{45} (\bibinfo{year}{1956}).

\bibitem[{\citenamefont{Yosida}(1957)}]{Yosida1957}
\bibinfo{author}{\bibfnamefont{K.}~\bibnamefont{Yosida}},
  \bibinfo{journal}{Phys. Rev.} \textbf{\bibinfo{volume}{106}},
  \bibinfo{pages}{893} (\bibinfo{year}{1957}).

\bibitem[{\citenamefont{Yambe and Hayami}(2022)}]{Yambe_PhysRevB.106.174437}
\bibinfo{author}{\bibfnamefont{R.}~\bibnamefont{Yambe}} \bibnamefont{and}
  \bibinfo{author}{\bibfnamefont{S.}~\bibnamefont{Hayami}},
  \bibinfo{journal}{Phys. Rev. B} \textbf{\bibinfo{volume}{106}},
  \bibinfo{pages}{174437} (\bibinfo{year}{2022}).

\bibitem[{\citenamefont{Akagi et~al.}(2012)\citenamefont{Akagi, Udagawa, and
  Motome}}]{Akagi_PhysRevLett.108.096401}
\bibinfo{author}{\bibfnamefont{Y.}~\bibnamefont{Akagi}},
  \bibinfo{author}{\bibfnamefont{M.}~\bibnamefont{Udagawa}}, \bibnamefont{and}
  \bibinfo{author}{\bibfnamefont{Y.}~\bibnamefont{Motome}},
  \bibinfo{journal}{Phys. Rev. Lett.} \textbf{\bibinfo{volume}{108}},
  \bibinfo{pages}{096401} (\bibinfo{year}{2012}).

\bibitem[{\citenamefont{Hayami and Motome}(2014)}]{Hayami_PhysRevB.90.060402}
\bibinfo{author}{\bibfnamefont{S.}~\bibnamefont{Hayami}} \bibnamefont{and}
  \bibinfo{author}{\bibfnamefont{Y.}~\bibnamefont{Motome}},
  \bibinfo{journal}{Phys. Rev. B} \textbf{\bibinfo{volume}{90}},
  \bibinfo{pages}{060402(R)} (\bibinfo{year}{2014}).

\bibitem[{\citenamefont{Hayami and
  Motome}(2018)}]{Hayami_PhysRevLett.121.137202}
\bibinfo{author}{\bibfnamefont{S.}~\bibnamefont{Hayami}} \bibnamefont{and}
  \bibinfo{author}{\bibfnamefont{Y.}~\bibnamefont{Motome}},
  \bibinfo{journal}{Phys. Rev. Lett.} \textbf{\bibinfo{volume}{121}},
  \bibinfo{pages}{137202} (\bibinfo{year}{2018}).

\bibitem[{\citenamefont{Hayami}(2022{\natexlab{a}})}]{hayami2022square}
\bibinfo{author}{\bibfnamefont{S.}~\bibnamefont{Hayami}}, \bibinfo{journal}{J.
  Phys.: Condens. Matter} \textbf{\bibinfo{volume}{34}},
  \bibinfo{pages}{365802} (\bibinfo{year}{2022}{\natexlab{a}}).

\bibitem[{\citenamefont{Hayami and
  Motome}(2021{\natexlab{b}})}]{Hayami_PhysRevB.103.024439}
\bibinfo{author}{\bibfnamefont{S.}~\bibnamefont{Hayami}} \bibnamefont{and}
  \bibinfo{author}{\bibfnamefont{Y.}~\bibnamefont{Motome}},
  \bibinfo{journal}{Phys. Rev. B} \textbf{\bibinfo{volume}{103}},
  \bibinfo{pages}{024439} (\bibinfo{year}{2021}{\natexlab{b}}).

\bibitem[{\citenamefont{Hayami and
  Kato}(2023{\natexlab{b}})}]{hayami2023widely}
\bibinfo{author}{\bibfnamefont{S.}~\bibnamefont{Hayami}} \bibnamefont{and}
  \bibinfo{author}{\bibfnamefont{Y.}~\bibnamefont{Kato}}, \bibinfo{journal}{J.
  Magn. Magn. Mater.} \textbf{\bibinfo{volume}{571}}, \bibinfo{pages}{170547}
  (\bibinfo{year}{2023}{\natexlab{b}}).

\bibitem[{\citenamefont{Wang et~al.}(2021)\citenamefont{Wang, Su, Lin, and
  Batista}}]{Wang_PhysRevB.103.104408}
\bibinfo{author}{\bibfnamefont{Z.}~\bibnamefont{Wang}},
  \bibinfo{author}{\bibfnamefont{Y.}~\bibnamefont{Su}},
  \bibinfo{author}{\bibfnamefont{S.-Z.} \bibnamefont{Lin}}, \bibnamefont{and}
  \bibinfo{author}{\bibfnamefont{C.~D.} \bibnamefont{Batista}},
  \bibinfo{journal}{Phys. Rev. B} \textbf{\bibinfo{volume}{103}},
  \bibinfo{pages}{104408} (\bibinfo{year}{2021}).

\bibitem[{\citenamefont{Hayami}(2022{\natexlab{b}})}]{Hayami_PhysRevB.105.174437}
\bibinfo{author}{\bibfnamefont{S.}~\bibnamefont{Hayami}},
  \bibinfo{journal}{Phys. Rev. B} \textbf{\bibinfo{volume}{105}},
  \bibinfo{pages}{174437} (\bibinfo{year}{2022}{\natexlab{b}}).

\bibitem[{\citenamefont{Utesov}(2021)}]{Utesov_PhysRevB.103.064414}
\bibinfo{author}{\bibfnamefont{O.~I.} \bibnamefont{Utesov}},
  \bibinfo{journal}{Phys. Rev. B} \textbf{\bibinfo{volume}{103}},
  \bibinfo{pages}{064414} (\bibinfo{year}{2021}).

\bibitem[{\citenamefont{Hayami}(2022{\natexlab{c}})}]{hayami2022multiple}
\bibinfo{author}{\bibfnamefont{S.}~\bibnamefont{Hayami}}, \bibinfo{journal}{J.
  Phys. Soc. Jpn.} \textbf{\bibinfo{volume}{91}}, \bibinfo{pages}{023705}
  (\bibinfo{year}{2022}{\natexlab{c}}).

\bibitem[{\citenamefont{Hayami and Yambe}(2022)}]{Hayami_PhysRevB.105.104428}
\bibinfo{author}{\bibfnamefont{S.}~\bibnamefont{Hayami}} \bibnamefont{and}
  \bibinfo{author}{\bibfnamefont{R.}~\bibnamefont{Yambe}},
  \bibinfo{journal}{Phys. Rev. B} \textbf{\bibinfo{volume}{105}},
  \bibinfo{pages}{104428} (\bibinfo{year}{2022}).

\bibitem[{\citenamefont{Takagi et~al.}(2022)\citenamefont{Takagi, Matsuyama,
  Ukleev, Yu, White, Francoual, Mardegan, Hayami, Saito, Kaneko
  et~al.}}]{takagi2022square}
\bibinfo{author}{\bibfnamefont{R.}~\bibnamefont{Takagi}},
  \bibinfo{author}{\bibfnamefont{N.}~\bibnamefont{Matsuyama}},
  \bibinfo{author}{\bibfnamefont{V.}~\bibnamefont{Ukleev}},
  \bibinfo{author}{\bibfnamefont{L.}~\bibnamefont{Yu}},
  \bibinfo{author}{\bibfnamefont{J.~S.} \bibnamefont{White}},
  \bibinfo{author}{\bibfnamefont{S.}~\bibnamefont{Francoual}},
  \bibinfo{author}{\bibfnamefont{J.~R.~L.} \bibnamefont{Mardegan}},
  \bibinfo{author}{\bibfnamefont{S.}~\bibnamefont{Hayami}},
  \bibinfo{author}{\bibfnamefont{H.}~\bibnamefont{Saito}},
  \bibinfo{author}{\bibfnamefont{K.}~\bibnamefont{Kaneko}},
  \bibnamefont{et~al.}, \bibinfo{journal}{Nat. Commun.}
  \textbf{\bibinfo{volume}{13}}, \bibinfo{pages}{1472} (\bibinfo{year}{2022}).

\bibitem[{\citenamefont{Hayami}(2023)}]{hayami2023orthorhombic}
\bibinfo{author}{\bibfnamefont{S.}~\bibnamefont{Hayami}}, \bibinfo{journal}{J.
  Phys.: Mater.} \textbf{\bibinfo{volume}{6}}, \bibinfo{pages}{014006}
  (\bibinfo{year}{2023}).

\bibitem[{\citenamefont{Agterberg and
  Yunoki}(2000)}]{Agterberg_PhysRevB.62.13816}
\bibinfo{author}{\bibfnamefont{D.~F.} \bibnamefont{Agterberg}}
  \bibnamefont{and} \bibinfo{author}{\bibfnamefont{S.}~\bibnamefont{Yunoki}},
  \bibinfo{journal}{Phys. Rev. B} \textbf{\bibinfo{volume}{62}},
  \bibinfo{pages}{13816} (\bibinfo{year}{2000}).

\bibitem[{\citenamefont{Martin and
  Batista}(2008)}]{Martin_PhysRevLett.101.156402}
\bibinfo{author}{\bibfnamefont{I.}~\bibnamefont{Martin}} \bibnamefont{and}
  \bibinfo{author}{\bibfnamefont{C.~D.} \bibnamefont{Batista}},
  \bibinfo{journal}{Phys. Rev. Lett.} \textbf{\bibinfo{volume}{101}},
  \bibinfo{pages}{156402} (\bibinfo{year}{2008}).

\bibitem[{\citenamefont{Hayami and
  Yambe}(2020)}]{Hayami_doi:10.7566/JPSJ.89.103702}
\bibinfo{author}{\bibfnamefont{S.}~\bibnamefont{Hayami}} \bibnamefont{and}
  \bibinfo{author}{\bibfnamefont{R.}~\bibnamefont{Yambe}}, \bibinfo{journal}{J.
  Phys. Soc. Jpn.} \textbf{\bibinfo{volume}{89}}, \bibinfo{pages}{103702}
  (\bibinfo{year}{2020}).

\bibitem[{\citenamefont{Hayami and
  Motome}(2021{\natexlab{c}})}]{Hayami_PhysRevB.104.144404}
\bibinfo{author}{\bibfnamefont{S.}~\bibnamefont{Hayami}} \bibnamefont{and}
  \bibinfo{author}{\bibfnamefont{Y.}~\bibnamefont{Motome}},
  \bibinfo{journal}{Phys. Rev. B} \textbf{\bibinfo{volume}{104}},
  \bibinfo{pages}{144404} (\bibinfo{year}{2021}{\natexlab{c}}).

\bibitem[{\citenamefont{Zener}(1951)}]{Zener_PhysRev.82.403}
\bibinfo{author}{\bibfnamefont{C.}~\bibnamefont{Zener}},
  \bibinfo{journal}{Phys. Rev.} \textbf{\bibinfo{volume}{82}},
  \bibinfo{pages}{403} (\bibinfo{year}{1951}).

\bibitem[{\citenamefont{Anderson and
  Hasegawa}(1955)}]{anderson1955considerations}
\bibinfo{author}{\bibfnamefont{P.~W.} \bibnamefont{Anderson}} \bibnamefont{and}
  \bibinfo{author}{\bibfnamefont{H.}~\bibnamefont{Hasegawa}},
  \bibinfo{journal}{Phys. Rev.} \textbf{\bibinfo{volume}{100}},
  \bibinfo{pages}{675} (\bibinfo{year}{1955}).

\end{thebibliography}
\end{document}